\documentclass[a4paper,11pt]{article}
\pdfoutput=1 

\usepackage{jcappub} 

\usepackage[T1]{fontenc} 
\usepackage{color}

\title{\boldmath Gravitational lensing by magnetized compact object in the presence of plasma}

\author[a]{Bobur Turimov,}
\author[a,b]{Bobomurat Ahmedov,}
\author[a,c,1]{Ahmadjon~Abdujabbarov,\note{Corresponding author.}}
\author[c,d]{and Cosimo Bambi}

\affiliation[a]{Ulugh Beg Astronomical Institute,\\ Astronomicheskaya 33,
Tashkent  100052, Uzbekistan}
\affiliation[b]{National University of Uzbekistan, Tashkent 100174, Uzbekistan}
\affiliation[c]{Center for Field Theory and Particle Physics and Department of Physics, Fudan University, 200433 Shanghai, China}
\affiliation[d]{Theoretical Astrophysics, Eberhard-Karls Universit\"{a}t T\"{u}bingen, 72076 T\"{u}bingen, Germany}

\emailAdd{bturimov@astrin.uz}
\emailAdd{ahmedov@astrin.uz}
\emailAdd{ahmadjon@fudan.edu.cn}
\emailAdd{bambi@fudan.edu.cn}

\abstract{
We study the gravitational lensing in the weak field approximation assuming the presence of a plasma and of a magnetic field around a compact gravitational source. The external magnetic field causes the split of the image, as the counterpart of the Zeeman effect. The magnetic field affects the magnification of images, creating additional components. We also study the time delay of an electromagnetic signal due to the geometry and the gravitational field around the source. We show that the time delay strongly depends on the plasma parameters. Lastly, we consider the effects of the presence of an inhomogeneous plasma on the gravitational lensing.}

\begin{document}
\maketitle
\flushbottom

\section{Introduction}
\label{sec:intro}

Light deflection and lensing in curved spacetime due to the presence of matter and/or energy density is one of distinguished features of general relativity. The gravitational lensing in the weak field approximation has been studied by many authors. The first studies on microlensing are in Refs.~\cite{Paczynski86a, Alcock93, Aubourg93, Udalski93, Paczynski96}. A review article on this subject is Ref.~\cite{Perlick04}. Strong gravitational lensing and the angular sizes and magnification factors for relativistic rings formed by photons were studied in~\cite{Tsupko09}. The gravitational lensing by wormholes was considered in~\cite{Nandi06}. The authors of Ref.~\cite{Schee15} studied the gravitational lensing and the ghost images in the regular Bardeen no-horizon spacetimes.

The interaction of the photons with the plasma surrounding the gravitational lens in the presence of a strong magnetic field is particularly interesting from both the theoretical and the observational point of view. According to the no-hair theorem, astrophysical black holes do not posses their own magnetic field. Nevertheless, an external magnetic field can be generated by the current of the surrounding plasma or (in binary systems) by the companion star, if the latter is a neutron star or a magnetar with a strong magnetic field. The electromagnetic field configuration in the vicinity of a black hole immersed in an external magnetic field was first considered by Wald~\cite{Wald74}. After that seminal work, many other authors have studied the properties of the spacetime around a black hole immersed in an external magnetic field~\cite{Aliev86, Aliev89, Aliev02, Aliev2004, Frolov12, Frolov10, Toshmatov15d, Stuchlik14a, Abdujabbarov08, Abdujabbarov14, Abdujabbarov10, Abdujabbarov11a, Abdujabbarov11, Abdujabbarov13a, Abdujabbarov13b}. The magnetic field of a current loop around a black hole was considered in~\cite{Petterson74}.  The charged-fluid toroidal structures surrounding a static charged black hole in an asymptotically uniform magnetic field was studied in~\cite{Kovar14}.
The role of gravitational lensing on the study of the distribution of stars in the Milky Way and on the study of dark matter and dark energy on very large scales is discussed in~\cite{DePaolis16}. An overview of the problems of the plasma influence on the effects of gravitational lensing is reported in~\cite{Kogan17}. Optical effects related to Keplerian discs orbiting Kehagias-Sfetsos naked singularities were discussed in~\cite{Stuchlik14}. In Ref.~\cite{Schee09}, the optical phenomena in the field of braneworld Kerr black holes were studied in detail.

Gravitational lensing by a rotating massive object in a plasma has been considered by different authors~\cite{Perlick04a,Perlick04,Grenzebach2014,Rogers15,Grenzebach15,Tsupko10,Morozova13,Tsupko14,Perlick15,Perlick17}. The recent study in Ref.~\cite{Abdujabbarov17} is devoted to gravitational lensing by different types of regular black holes in the presence of plasma. Various optical properties of black holes in the presence of plasma, like the black hole shadow, were studied in Refs.~\cite{Abdujabbarov13c, Atamurotov13b, Atamurotov13, Atamurotov15a, Abdujabbarov15, Abdujabbarov16b, Abdujabbarov16a, Hakimov16, Abdujabbarov17b}.

The aim of the present work is to study the gravitational lensing in a weak gravitational field in the presence of plasma and magnetic field. The paper is organized as follows. In Sect.~\ref{theory}, we present the master equations for the description of the plasma around the black hole in the presence of a magnetic field and for the description of the photon motion. Sect.~\ref{observation} is devoted to the study of the observational consequences of the gravitational lensing in the presence of plasma and magnetic field. The summary of the results are reported in Sect.~\ref{conclusion}.

{Throughout the paper we employ the convention of a metric with signature ($-, +, +, +$). We use units in which $G=c=1$, but we restore $G$ and $c$ when we have to compare our findings with observational data. Greek indices run from $0$ to $3$, while Latin indices run from $1$ to $3$.}

\section{ Theoretical framework \label{theory}}

\subsection{Uniform magnetic field in the vicinity of a black hole}

{In this subsection, we will briefly discuss 
the electromagnetic field in the vicinity 
of a static compact object. We will also estimate the 
value of magnetic fields for supermassive and 
stellar-mass black holes, respectively.}
{ In order to study the 
propagation of the photons (electromagnetic waves) 
through a magnetized plasma in some arbitrary 
space-time we will consider magnetochydrodynamic equations
which is written as~\cite{Breuer80,Breuer81,Breuer81a} 
\begin{eqnarray}
&&
\label{ME1}
\nabla_{[\alpha} F_{\mu\nu]} = 0\ ,
\\
&&
\label{ME2}
\nabla_{\alpha} F^{\alpha\beta} = J^\beta\ ,
\\
&&
V^\alpha \nabla_\alpha V^\beta = \frac{q}{m} F_\alpha^\beta V^{\alpha}\ ,
\\
&&
\nabla_\alpha (N_qV^\alpha) = 0\ ,
\\
&&
V^\alpha V_\alpha = -1\ ,
\end{eqnarray}
where $F_{\alpha\beta} = \partial_\alpha A_\beta - \partial_\beta A_\alpha$ 
is the electromagnetic field tensor with the 
vector potential $A_\alpha$ of the electromagnetic field,
$J^\alpha$ is the electric current of the electrons and ions and,
$q$ and $m$ are the their charge and mass, respectively.
$N_q$ and $V^{\alpha}$ are the number density and four-velocity 
of the charged particles.  }

{ The fundamental equation for the small 
perturbation of the vector potential $\hat{A}_\alpha $ is given by ~\cite{Breuer80,Breuer81,Breuer81a} 
\begin{eqnarray}\label{funEq}
D^{\alpha\beta} \hat{A}_\beta = \left[h^{\alpha\mu}V^\nu
\nabla_\nu(\nabla_\mu^\beta-\delta_\mu^\beta 
\nabla_\lambda^\lambda) 
\right.
\\\nonumber
\left.
+ (\omega^{\alpha\mu} +
\omega_L^{\alpha\mu} +\Theta^{\alpha\mu}+\Theta h^{\alpha\mu}
+ \frac{q}{m} E^\alpha V^\mu )(\nabla_\mu^\beta-\delta_\mu^\beta 
\nabla_\lambda^\lambda)
\right.
\\\nonumber
\left.
+\omega_p^2 (h^{\alpha\beta} V^\lambda\nabla_\lambda+\Theta^{\alpha\beta}-\omega^{\alpha\beta})
\right]\hat{A}_\beta = 0 \ ,
\end{eqnarray}
where $\omega_p = (4\pi N_q q^2/m)^{1/2}$ is the 
plasma frequency and $\omega_{\rm L}$ is Larmor 
frequency which defined as
\begin{equation}
\label{Larmor}
\omega_{\rm L} = \sqrt{\omega_{\rm L \, \mu}\omega_{\rm L}^{\mu}}\ , 
\quad \omega_{\rm L}^{\mu} = \frac{q}{2m}\eta^{\mu\nu\alpha\beta}V_\nu 
B_{\alpha\beta} \ ,
\end{equation}
with
$
B_{\alpha\beta} = h_{\alpha}
^{\mu} h_\beta^\nu F_{\mu\nu} 
$, $h_\beta^\alpha = \delta_\beta^\alpha + V^\alpha V_\beta$ 
and $\eta_{\mu\nu\alpha\beta}$ is the 
Levi-Civitta symbol in four dimensional curved space
and other definitions are
$$
E^\alpha = F^{\alpha}_\beta V^\beta, \quad
\nabla_\beta V^\alpha = \omega_\beta^\alpha + \Theta_\beta^\alpha
-V_\beta V^\gamma\nabla_\gamma V^\alpha \ ,
$$
$$
\omega_{\alpha\beta} = -\omega_{\beta\alpha}= -\frac{q}{m}B_{\alpha\beta}\ , \quad 
\Theta_{\alpha\beta} = \Theta_{\beta\alpha}, \quad \Theta = \nabla_\alpha V^\alpha  \ .
$$
}

Let us consider the spacetime of a static black hole with total mass $M$. In spherical  coordinate $(t,r,\theta,\phi)$, the line element is 
\begin{eqnarray}\label{metric}
ds^2 &=& -N^2dt^2+
\frac{1}{N^2}dr^2 
+ 
r^2(d\theta^2 + \sin^2\theta d\phi^2)\ ,
\end{eqnarray}
where the lapse function has a form $N^2 = 1- 2M/r$.

{In Ref.~\cite{Petterson74}, the author 
considered a current loop around a static black hole. 
He solved the general relativistic Maxwell equations
(\ref{ME1}) and (\ref{ME2})  
for the vector potential $A_\alpha$ in the case of a 
current loop surrounding a black hole at the distance $R$. 
It was shown that the vector potential in both interior 
and exterior regions of the current can be expressed 
in the following dipolar form~\cite{Petterson74,preti}}
%
\begin{eqnarray}\label{Af}
A_{\alpha} = 
-\frac{3}{8}\delta_{\alpha}^\phi\frac{\mu r^2\sin^2\theta}{M^3} 
\left\{
\begin{array}{lcr}
g(R)
   \qquad 2M\leq r \leq R\ ,
   \\
   \\
   g(r)
   ,\qquad r \geq R \ ,
   \end{array}
\right.
\end{eqnarray}
with
$$
g(r) = \ln \left(1-\frac{2M}{r}\right)+
\frac{2M}{r}\left(1+\frac{M}{r}\right) \ ,
$$ 
where $\mu$ is the magnetic dipole moment and it 
can be written in terms of the electric current $I$ as
\begin{eqnarray}
\mu = \pi R^2 N_R I\ ,
\end{eqnarray}
where subscript $R$ denotes the function at $r=R$.

\textbf{Exterior magnetic field outside loop --}
The expression (\ref{Af}) for the 
vector potential of the electromagnetic field outside 
the current loop ($r>R$) allows one to find the 
components of the exterior magnetic field~\cite{Petterson74}%
\begin{eqnarray}
\label{Br}
B^{\hat r}(r,\theta) &=& -\frac{3\mu}{4M^3}
\left[
\ln N^2+\frac{2M}{r}\left(1+\frac{M}{r}\right)\right]\cos\theta\ ,
\\
B^{\hat\theta}(r,\theta) &=& \frac{3\mu N}{4M^2r}
\left[\frac{r}{M}\ln N^2+\frac{1}{N^2}+1\right]\sin\theta\ .
\label{Bt}
\end{eqnarray}
{Neglecting higher order terms in $M/r$ and $M/R$ 
one can estimate the magnetic field as}
%
\begin{eqnarray}
\lim_{M/r\to 0, M/R\to 0}B^{\hat r}(r,\theta) &=& 
\frac{2\mu_0}{r^3}\cos\theta\ ,
\\
\lim_{M/r\to 0, M/R\to 0}B^{\hat\theta}(r,\theta) &=& 
\frac{\mu_0 }{r^3}\sin\theta\ ,
\end{eqnarray}
{where $\mu_0 (\mu_0= \pi R^2\,I)$ 
is the Newtonian value of the
magnetic dipole moment.} 

{\textbf{ Exterior uniform magnetic field within loop --}
From Eq.~(\ref{Af}), we can easily see that inside the current loop ($2M<r<R$) the vector potential of the electromagnetic field can be expressed in terms of a uniform magnetic field B as}
\begin{eqnarray}\label{Aunif}
A_\phi &=& \frac{1}{2} B\,r^2\sin^2\theta \ ,
\end{eqnarray}
where the uniform magnetic field has the form 
\begin{eqnarray}
B = -\frac{3\mu }{4M^3}\left[\ln N^2_R + 
\frac{2M}{R}\left(1+\frac{M}{R}\right)\right] \ ,
\end{eqnarray}
{and is oriented along the $z$-axes. If we use the expression for the vector potential in Eq.~(\ref{Aunif}), we find the components of the interior magnetic field}
\begin{eqnarray}
B^{\hat r} &=& B\cos\theta\ ,
\\
B^{\hat \theta} &=& BN\sin\theta \ .
\end{eqnarray}
{The total magnetic field is $B_{\rm T} = (B^{\hat r 2}+B^{\hat\theta 2})^{1/2}$. In the limit of weak gravitational field, we can estimate the magnetic field as $B_{\rm T} \simeq B$ and we get} 
\begin{eqnarray}\label{B}
B_{\rm T} \simeq B =  -\frac{3\mu }{4M^3}
\left[\ln N^2_R+ 
\frac{2M}{R}\left(1+\frac{M}{R}\right)\right] \ . 
\end{eqnarray}

{\textbf{ Quasi-uniform magnetic field case --} Let us now assume that the plasma is axially symmetrically distributed around a Schwarzschild black hole. An electromagnetic wave (photon) does not interact with the magnetic field, because in the vacuum a magnetic field does not affect the propagation of light rays. This means that it does not matter whether the magnetic field has a dipolar (or multipolar) structure outside the plasma. However, when the light ray propagates in the magnetized plasma (medium) it can be in a resonance state due to the cyclotron frequency of charged particles. We thus use the expression in Eq.~(\ref{B}) for the magnetic field in the interior region of the current loop. For simplicity, we assume that the magnetic field is uniform in the vicinity of the black hole, in particular near the equatorial plane ($\theta \simeq \pi/2$).  
} 

In order to estimate the magnetic field strength in the interior region of the loop, we consider the zone near the horizon of the black hole ($2M < r < R$). Let us assume that the electric current loop 
is located at the radial coordinate $R = 6 M$, corresponding to the radius of the innermost stable circular orbit (ISCO) for a test-particle around a Schwarzschild black hole. We can now evaluate the magnetic field in the expression in Eq.~(\ref{B}) as
\begin{eqnarray}\label{B0}
B \simeq \frac{I}{M} \to \frac{I}{GM/c^2}\ ,
\end{eqnarray}
which depends on the value of the current $I$ and  mass of the black hole $M$. We can define the electric current $I$ as
\begin{equation}\label{Current}
I = e\,n_j v_j\, S \ ,
\end{equation}
where $e$ is the electric charge of an electron, and $n_j$ and $v_j$ are, respectively, the density and the velocity of the charged particles in plasma. The subscript $"j"$ in Eq.~(\ref{Current}) refers to the type of particle. $S = \pi h l $ is the elliptic cross section of the accretion disc with the height $h$ and width~$l$.

Using Eqs.~(\ref{B0}) and (\ref{Current}) and the following input values
$$
n_e \sim 10^{5} \, {\rm cm}^{-3}
\ , \ \ 
h \sim 10^{4}\,  {\rm cm}
\ , \ \ 
l \sim 4 M\ , 
$$
we get an estimate for the typical magnetic field strength 
around stellar-mass black holes 
\begin{eqnarray}
B \simeq 6.5 \left(\frac{n_e}{10^5 
{\rm cm}^{-3}}\right)\left(\frac{h}{10^4 {\rm cm}}\right)
\left(\frac{l}{4M}\right)
\left(\frac{M}{M_\odot}\right)^{-1} 10^8 \, {\rm G}
 \ .\nonumber\\
\end{eqnarray}

Similarly, using the input values 
$$
n_e \sim 10^{4} cm^{-3}
\ , \ \  h \sim 10^{5} cm
\ , \ \  l \sim 4 \times 10^{-5}  M , 
$$
we get an estimation for the magnetic field strength around supermassive black holes 
\begin{eqnarray}
B &\simeq & 4.3 \left(\frac{n_e}{10^4 
{\rm cm}^{-3}}\right)\left(\frac{h}{10^5 {\rm cm}}\right)
\left(\frac{l}{4\cdot 10^{-5}M}\right)\left(\frac{M}{10^6M_\odot}\right)^{-1} 
10^4 \, {\rm G} \ .
\end{eqnarray}
Note that in both cases the particle is mildly relativistic $v \simeq 0.4 c$.

\subsection{Photon motion in the plasma surrounding a black hole}

In this subsection, we will consider the photon motion around a static black hole taking into account that the compact object is surrounded by a plasma and there is a non-vanishing magnetic field. At large distances from the black hole, the spacetime geometry tends to be flat and thus photons move along straight lines. The photons approaching the central object deviate from a straight line path.
In order to study the photon trajectory, we consider the following set of the differential equations~\cite{Synge60}:
\begin{eqnarray}\label{Aeq_of_mot}
\frac{dx^\mu}{d\lambda} = \frac{\partial H}{\partial
p_\mu}\ ,\qquad 
\frac{dp_\mu }{d\lambda} = -\frac{\partial H}{\partial x^\mu}\ ,
\end{eqnarray}
where $\lambda$ is an affine parameter depending on proper time $\tau$. $H$ is the Hamiltonian of the photon and can be written as~\cite{Synge60}
\begin{eqnarray}\label{Hamiltinian}
H(x^\mu ,p_\mu)=\frac{1}{2}\left[g^{\mu\nu}p_\mu p_\nu -(n^2-1)\left(p_{\mu} V^{\mu}\right)^2\right]=0\ . \nonumber\\
\end{eqnarray}
In Eq.~(\ref{Hamiltinian}), $n$ is the refractive index of the medium, $p^{\mu}$ is the 4-momentum of the photon, and {$V^{\mu}$ is the 4-velocity of the medium.} 
According to Ref.~\cite{Synge60}, we have to take into consideration the following relation between the momentum and the 4-velocity of the photon in the medium:
\begin{equation}\label{relation}
p_{\mu} V^{\mu}=-\frac{\hbar\omega(x^i)}{c}\ ,
\end{equation}
were $\omega (x^i)$ is the photon frequency in the medium, and $\hbar$ and $c$ are, respectively, the Planck constant and the speed of light in the vacuum.

Assuming that the photon is moving along the $z$-axis in flat spacetime, the 4-momentum $p^{\mu}$ can be written as~\cite{Kogan10}
\begin{eqnarray}
\label{Eq2}
p^{\mu} &=& \frac{\hbar\omega}{c}\left(1,0,0,n\right)\ ,
\quad
p_{\mu} = \frac{\hbar\omega}{c}\left(-1,0,0,n\right)\ .
\end{eqnarray}

Let us now consider the weak field limit. The covariant and 
contravariant components of the metric tensor can be written as
\begin{equation} 
g_{\mu\nu} = \eta_{\mu\nu} + h_{\mu\nu}\ \ 
 {\rm and}\ \  g^{\mu\nu} = \eta^{\mu\nu} - h^{\mu\nu}\ , 
\end{equation}
where $\eta_{\mu\nu}$ is the metric tensor in Minkowski space, $h_{\mu\nu}$ is a small perturbation, and the following conditions hold
\begin{equation}
\eta_{\mu\nu} = \eta^{\mu\nu}\ ,\ \  
h_{\mu\nu} = h^{\mu\nu} \ \ {\rm  and} \ \ h_{\mu\nu} h^{\mu\nu} \to 0.
\end{equation}

In the presence of a weak inhomogeneous plasma and a weak gravitational field, the photon equations of motion can be rewritten as
\begin{eqnarray}\label{Eq3}
\frac{dz}{d\lambda}&=&\frac{n\hbar\omega}{c}\ ,
\\
\frac{dp_{\mu}}{dz}&=&\frac{1}{2}\frac{n\hbar\omega}{c}
\left(h_{zz ,\mu} +
\frac{1}{n^2}h_{tt ,\mu}
-\frac{\omega_{e,\mu}^2}{n^2\,\omega^2}\right)\ ,
\label{Eq4}
\end{eqnarray}
where $\omega_e$ is the frequency of the electron plasma and is defined as
\begin{equation}\label{plasmfreq}
\omega^2_e(x)=\frac{4\pi e^2 N_e(x)}{m_e}\ ,
\end{equation}
$N_e=N_e(x^{i})$ is the electron concentration with respect to the coordinate in the plasma, and $m_e$ is the electron mass. 

The deflection angle is defined as the difference between the directions of the incoming and of the outgoing light rays. Following Ref.~\cite{Kogan10}, we can write the expression of the deflection angle of the light ray in the plane perpendicular to the $z$-axis as
\begin{eqnarray}
\hat{\alpha}_i &=& 
{e}_{i\ {\rm in}}-{e}_{i\ {\rm out}}\ ,
\end{eqnarray}
where ${e}_i$ is the unit vector along the vector $p_i$, i.e. ${e}_i = {p}_i/p$, and $p=\sqrt{p_x^2+p_y^2+p_z^2} = p_z = n \hbar\omega/c$. Employing Eqs.~(\ref{Eq3}) and (\ref{Eq4}), 
we obtain the formula for the absolute value of the deflection angle
\begin{eqnarray}
\label{alphaz}
\alpha =\vert \hat{\bold \alpha}_k\vert = \frac{1}{2}\bigg\vert\int^{\infty}_{-\infty}
\frac{\partial}{\partial
x^{k}}\Bigg(h_{zz}+\frac{1}{n^2}h_{tt}
-\frac{\omega_{e}^2}{n^2\omega^2}\Bigg)dz\bigg\vert \ , \nonumber\\
\end{eqnarray}

For an inhomogeneous plasma, the refraction index $n$ 
depends not only on the frequency of the electron 
plasma $\omega_e$ but also on the magnetic field generated 
by the accretion disk of the black hole. 
From the fundamental equation (\ref{funEq}) one can obtain
dispersion relation as follows (See e.g., ~\cite{Breuer81,Breuer81a} )
\begin{eqnarray}\label{DisRel}
(\omega^2-k^2)\left[\omega^2\omega^2_{\rm L}(\omega^2-\omega^2_e-k^2)
+\omega^2_e({\bf\omega_{\rm L}\cdot k})^2\right] 
\\\nonumber
-\omega^2(\omega^2-\omega^2_e)(\omega^2-\omega^2_e-k^2)^2= 0\ ,
\end{eqnarray}
where $\bf k$ is the electromagnetic wave vector. After considering 
${\bf\omega_{\rm L}\cdot k} = \omega_{\rm L}k \cos\psi $. 
Taking the inhomogeneity of 
the plasma and the presence of a magnetic field into account, 
the refraction index can be written as 
\begin{equation}
\label{Arefrac} 
n^2 = n_\pm^2 = 1-\frac{\omega^2_e}{\omega^2}
-\frac{\omega^2_e}{\omega^2}\frac{\omega_{\rm L}}{\omega}
\,f_{\pm}\left(\omega_{\rm L},\omega_e\right)\ ,
\end{equation}
where $\psi$ is the angle between the magnetic field 
relative to the direction of the photon,
and the unknown function $f_{\pm}(\omega_{\rm L},\omega_e)$ is defined as
\begin{equation}
\label{f} 
f_{\pm}\left(\omega_{\rm L},\omega_e\right)
=
\frac{1}{2}\frac{\omega\omega_{\rm L}(\omega^2+(\omega^2-2\omega_e^2)
\cos^2\psi)\pm \omega^2\sqrt{4(\omega^2-\omega_e^2)^2
\cos^2\psi+\omega^2\omega_{\rm L}^2\sin^4\psi}}{\omega^2(\omega^2-
\omega_e^2-\omega_{\rm L}^2)+\omega_e^2\omega_{\rm L}^2\cos^2\psi} \ ,
\end{equation}
In the case when $\psi = 0 $
the expression (\ref{Arefrac}) for the refractive 
index takes the following form  
\begin{eqnarray}
n_\pm^2 =
1-\frac{\omega_e^2}{\omega^2}+\frac{\omega_e^2\omega_{\rm L}}
{\omega^2(\omega\mp\omega_{\rm L})} \ ,
\end{eqnarray} 
which is responsible for the case when photon comes to 
parallel to the magnetic field line,
while $\psi =\pi/2$ case when magnetic field line 
perpendicular to the direction of the photons and 
the refractive index can take a form   
\begin{eqnarray}
n_+^2 = 1-\frac{\omega_e^2}{\omega^2}
+\frac{\omega_e^2}{\omega^2}
\frac{\omega_{\rm L}^2}{\omega^2-\omega_e^2-\omega_{\rm L}^2} \ ,
\quad n_-^2 = 1-\frac{\omega_e^2}{\omega^2} \ ,
\end{eqnarray} 
From the equation (\ref{Arefrac}) one can easily see 
that absence of the magnetic field ($\omega_{\rm L} = 0$) 
the refractive index takes simple form $n^2 = 1-
\omega_e^2/\omega^2$ as in~\cite{Kogan10}.
Using the Eq.(\ref{Larmor}) and after doing some algebraic 
calculation one can obtain the explicit form of the 
Larmor frequency for the electron in the following form
\begin{eqnarray}
\label{LarmorF}
\omega_{\rm L}(r,\theta) &=& \frac{e}{m_e}\sqrt{\frac{1}{2}F_{\mu\nu}F^{\mu\nu}}
= \omega_{\rm c}\sqrt{1 -\frac{2M}{r}\sin^2\theta}\ ,
\end{eqnarray}
where $\omega_{\rm c}$ is the cyclotron frequency 
due to the uniform magnetic field $B$ and can be written as
\begin{equation}
\omega_{\rm c} = \frac{eB}{m_e c}\ ,
\end{equation}
which is same quantity as Larmor frequency
for the electron in the external uniform magnetic field.

We can estimate the typical cyclotron frequency for supermassive and stellar-mass black holes:  
\begin{itemize}
\item Supermassive black holes are located at the centre of galaxies and they have a mass in the range $M \sim 10^6-10^{10}\, M_{\odot}$. Typical values of $B$, $\omega_{\rm c}$, and $\lambda$ are~\cite{Piotrovich10,Baczko16,Fish11,Johnson15}:  
$$
B \sim 10^4 \,{\rm G},\quad \omega_{\rm c} \sim 30 \, 
{\rm GHz}, \quad \lambda \sim 6.3 \, {\rm cm} \, .
$$
Here $\omega_{\rm c}$ is at super high radio frequencies.
\item Stellar-mass black holes in the known X-ray binaries have a mass in the range $M \sim 3- 20\,M_{\odot}$. Typical values of $B$, $\omega_{\rm c}$, and $\lambda$ are~\cite{Donati05}:
$$
B \sim 10^8 \,{\rm G},\quad \omega_{\rm c} \sim 300\, 
{\rm THz}, \quad \lambda \sim 6.3\times 10^{-4} \, {\rm cm} .
$$
Now $\omega_{\rm c}$ is in the infrared spectrum.
\end{itemize}

Since very-long baseline interferometry observations are supposed to detect the radiation emitted by the accreting gas around the event horizon of the supermassive black holes Sgr A* and M87, it is important to study these effects in view of their near-future detectability.

\subsection{Polarization angle of the light in the medium}

In this subsection, we will consider the polarization angle of the light due to Faraday rotation in the presence of a magnetized plasma in the background of a static compact object. In Refs.~\cite{Narasimha08} and \cite{Sereno04}, the rotation angle of the polarization plane during the propagation of the light ray in the plasma is considered at the leading order of the magnetic field. Here we will use the following more general form of the polarization angle
\begin{eqnarray}\label{fi}
\Delta\varphi
&=& \int ds\, k\,(n_- - n_+)
\simeq  
\int ds \, k\,\frac{\omega_e^2\omega_{\rm L}}{2\omega^3}\left(f_+-f_-\right)
\\\nonumber
&=&  
\int ds \, k\,\frac{\omega_e^2\omega_{\rm L}}{2\omega}
\frac{\sqrt{4(\omega^2-\omega_e^2)^2
\cos^2\psi+\omega^2\omega_{\rm L}^2\sin^4\psi}}{\omega^2(\omega^2-
\omega_e^2-\omega_{\rm L}^2)+\omega_e^2\omega_{\rm L}^2\cos^2\psi}\ ,
\end{eqnarray}
where $k$ is the absolute value of the wave vector and 
can be written in terms of the frequency 
$k=|\bold{k}| = 2\pi/\lambda = \omega$. Then polarization angle 
in the expression (\ref{fi}) will take a form for the 
different values of the inclination angle 
\begin{eqnarray}
\Delta\varphi =
\int ds \,\omega_e^2\omega_{\rm L}\left\{
\begin{array}{lcr}
\frac{1}{\omega^2-\omega_{\rm L}^2},
   \qquad \psi = 0\ ,
   \\
   \\
\frac{\omega_{\rm L}}{2\omega(\omega^2-
\omega_e^2-\omega_{\rm L}^2)}
   ,\quad \psi = \pi/2 \ ,
   \end{array}
\right.
\end{eqnarray} 
By using the expression (\ref{fi}) for the polarization angle 
the rotation measure $RM$ can be calculated as
\begin{eqnarray}
RM 
&=& \frac{\Delta\varphi}{\lambda^2} 
=\frac{\omega^2}{4\pi^2}\, \Delta\varphi 
\\\nonumber
\end{eqnarray}
%

The scattering cross-sections of right-hand ``$+$'' and left-hand ``$-$'' polarized photons are different and can be calculated in the following way
\begin{eqnarray}\label{CS}
\sigma_\pm = \sigma_{\rm Th}
\left (1 \pm \frac{\omega_{\rm L}}{\omega}\cos\psi\right)\ , 
\end{eqnarray}
where $\sigma_{\rm Th} = (8\pi/3)(e^2/m_ec^2)^2$ 
is the classical Thomson cross section. 
According to Eq.~(\ref{CS}), the 
light ray from a magnetized plasma around a 
compact object is circularly polarized.

\subsection{Deflection angle around a static black hole}

This subsection is devoted to find the deflection angle of a photon moving in an inhomogeneous magnetized plasma in the spacetime of a static compact object. In the weak field approximation, the metric in Eq.~(\ref{metric}) can be rewritten in following form
\begin{eqnarray}\label{Wmetric}
ds^2 = ds_0^2 + \frac{2M}{r}\,dt^2 + \frac{2M}{r}\,dr^2\ ,
\end{eqnarray}
where $ds_0^2 = -dt^2+dx^2+dy^2+dz^2$ is the line element in flat space. The components of small perturbations of the metric tensor are $h_{\alpha\beta}$ and in the Cartesian frame are~\cite{Landau-Lifshitz2}
$$
h_{tt}=\frac{2M}{r}\ ,\quad h_{ij} =\frac{2M}{r}
\hat{n}_i\hat{n}_j\ , \quad h_{zz} =\frac{2M}{r}
\cos^2\theta ,
$$
where $\hat{n}_i$ is the component of the unit vector with the same direction as the radius vector $r_i = (x,y,z)$ and has the form $\hat n_i = (\cos\phi\sin\theta, \sin\phi\sin\theta, \cos\theta)$. 
Before calculating the deflection angle, it is useful to introduce 
the form of the plasma frequency as  $\omega_e(r) =\omega_0 (R_0/r)^h$, 
where $h$ and $R_0$ are constants and $\omega_0$ is the plasma 
frequency of the plasma at infinity. Using Eq.~(\ref{alphaz}), 
we get the expression for the deflection angle of a light ray 
passing near a magnetized static compact object: 
\begin{eqnarray}
\label{Dangle}
\alpha^{\pm} &=&
\frac{2M}{b}\left[1+\left(1-\frac{\omega^2_0}{\omega^2}
-\frac{\omega^2_0}{\omega^2}\frac{\omega_c}{\omega}
\,f_{\pm}\left(\omega_c,\omega_0\right)\right)^{-1} \right]
\\\nonumber
&-&
\frac{\omega_0^2}{\omega^2}\frac{\sqrt{\pi}\Gamma[(h+1)/2]}{\Gamma(h/2)}\left(\frac{R_0}{b}\right)^h
\left(1-\frac{\omega^2_0}{\omega^2}
-\frac{\omega^2_0}{\omega^2}\frac{\omega_c}{\omega}
\,f_{\pm}\left(\omega_c,\omega_0\right)\right)^{-1} + {\cal O}(M^2/b^2) \ ,
\end{eqnarray}
where $\Gamma(x)$ is the gamma function
$$
\Gamma(x)= \int_0^\infty t^{x-1}e^{-t}dt\ .
$$
In the case when $\omega \gg \omega_c$, we have
\begin{eqnarray}\label{Alfa}
\alpha^{\pm}
&\simeq &
\frac{2M}{b}\left(1+\frac{\omega^2}{\omega^2-\omega_0^2}
\pm 
\frac{\omega^3\omega_0^2 \omega_c\cos\psi}{\left(\omega^2-\omega_0^2\right)^2}\right)
\\\nonumber
&-&
\frac{\omega_0^2}{\omega^2}\frac{\sqrt{\pi}\Gamma[(h+1)/2]}{\Gamma(h/2)}\left(\frac{R_0}{b}\right)^h
\left(\frac{\omega^2}{\omega^2-\omega_0^2}
\pm 
\frac{\omega^3\omega_0^2 \omega_c\cos\psi}
{\left(\omega^2-\omega_0^2\right)^2}\right)\ .
\end{eqnarray}

It is worth noting that in Eq.~(\ref{Dangle}) there are four resonance states corresponding to the solutions of the equation 
$ \omega^3 - 2\,\omega_0^2\omega_c f_{\pm}
(\omega_c,\omega_0)-\omega\omega_0^2 = 0 $. 

\section{ Observational effects \label{observation}}

\begin{figure}
\begin{center}

\includegraphics[width=0.5\textwidth]{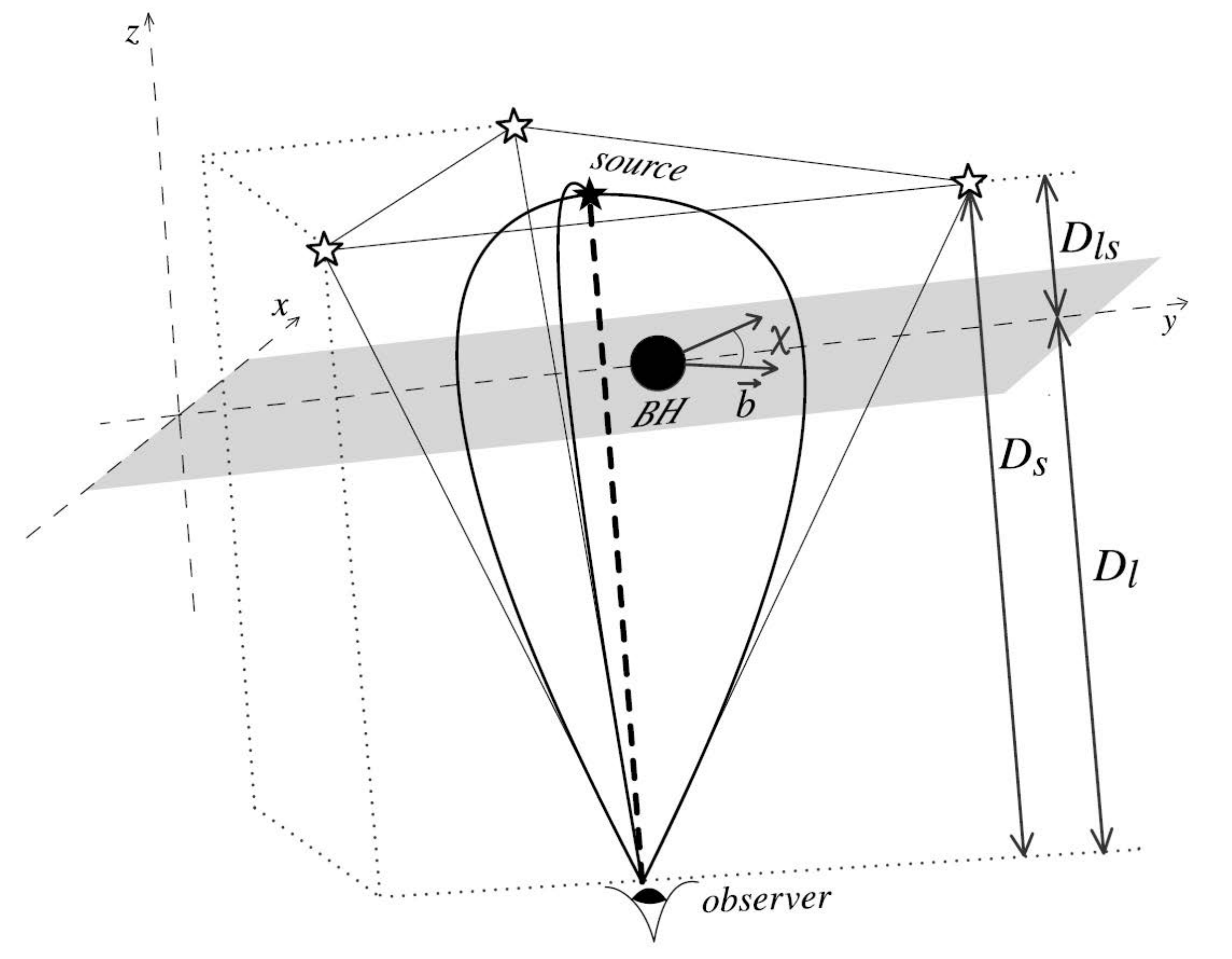}

\end{center}
\caption{The black hole $BH$ is between the $source$ and the $observer$. The light emitted by the $source$ and detected by the $observer$ is affected by the gravitational lensing of the $BH$. }
\label{GL}
\end{figure}

We can now study the observational consequences of gravitational lensing, namely the magnification of image sources, Einstein rings, etc. For this purpose, we can use the lens equation, which relates the angle $\beta$ of the real object from the observer-lens axis, the angle $\theta$ of the apparent image of the object from the observer-lens axis, and the deflection angle $\alpha$:
\begin{eqnarray}\label{Maineq}
D_{\rm s} \,\theta = D_{\rm s}\, \beta + 
D_{\rm ls}\,\alpha, 
\end{eqnarray}
where $D_{\rm s}$ is the distance between the observer and the source and $D_{\rm ls}$ is the distance between the lens and the source (see Fig.~\ref{GL}).
In the weak field approximation, we can use the relation $\alpha \sim 1/b $ in order to express the small angle $\theta =b/D_l$, where $D_{\rm l}$ is the distance between the observer and the lens, as shown in Fig~\ref{GL}. Hereafter, we will consider two cases: $i)$ homogeneous and $ii)$ inhomogeneous plasma around a gravitational source. We will always take the effects of magnetic field into account. 

\subsection{Homogeneous plasma}

In the case of a homogeneous magnetized plasma, the plasma frequency in Eq.~(\ref{plasmfreq}) is constant and in the weak field approximation the lens equation reduces to 
\begin{eqnarray}\label{beta}
\beta &=&\theta - 
\frac{\Theta_\pm^{2}}{\theta}\ ,
\end{eqnarray}
where
\begin{eqnarray}\label{Theta}
\Theta_{\pm} = \Theta_{\rm E}\sqrt{\frac{1}{2}
\left[ 1 + \left(1-\frac{\omega^2_0}{\omega^2}
-\frac{\omega^2_0}{\omega^2}\frac{\omega_c}{\omega}
\,f_{\pm}\left(\omega_c,\omega_0\right)\right)^{-1} \right]}\ ,
\end{eqnarray}
and $\Theta_{\rm E}$ is defined as
\begin{eqnarray}
\Theta_{\rm E} = \sqrt{\frac{4M\, D_{\rm ls}}
{D_{\rm l}\,D_{\rm s}}}\ .
\end{eqnarray}
%
here $\Theta_\pm$ is the Einstein ring splitted into two rings due to the magnetic Zeeman effect. In the absence of magnetic fields, corresponding to the case $\omega_c = 0 $, we can obtain the value of the unsplitted $\Theta$ in the plasma, as was done in~\cite{Kogan10,Morozova13,Abdujabbarov17}. In vacuum ($\omega_0=0$) $\Theta=\Theta_E$.

The split of the Einstein ring due to the magnetic field, or ``Zeeman effect'', expressed in Eq.~(\ref{Theta}) is schematically shown in Fig.~\ref{ER}. The upper plot of Fig.~\ref{ER} illustrates how the rings split into two. The lower plot shows how the angle changes due to the presence of the  plasma as well as of the magnetic field.

\begin{figure}
\includegraphics[width=0.42\textwidth]{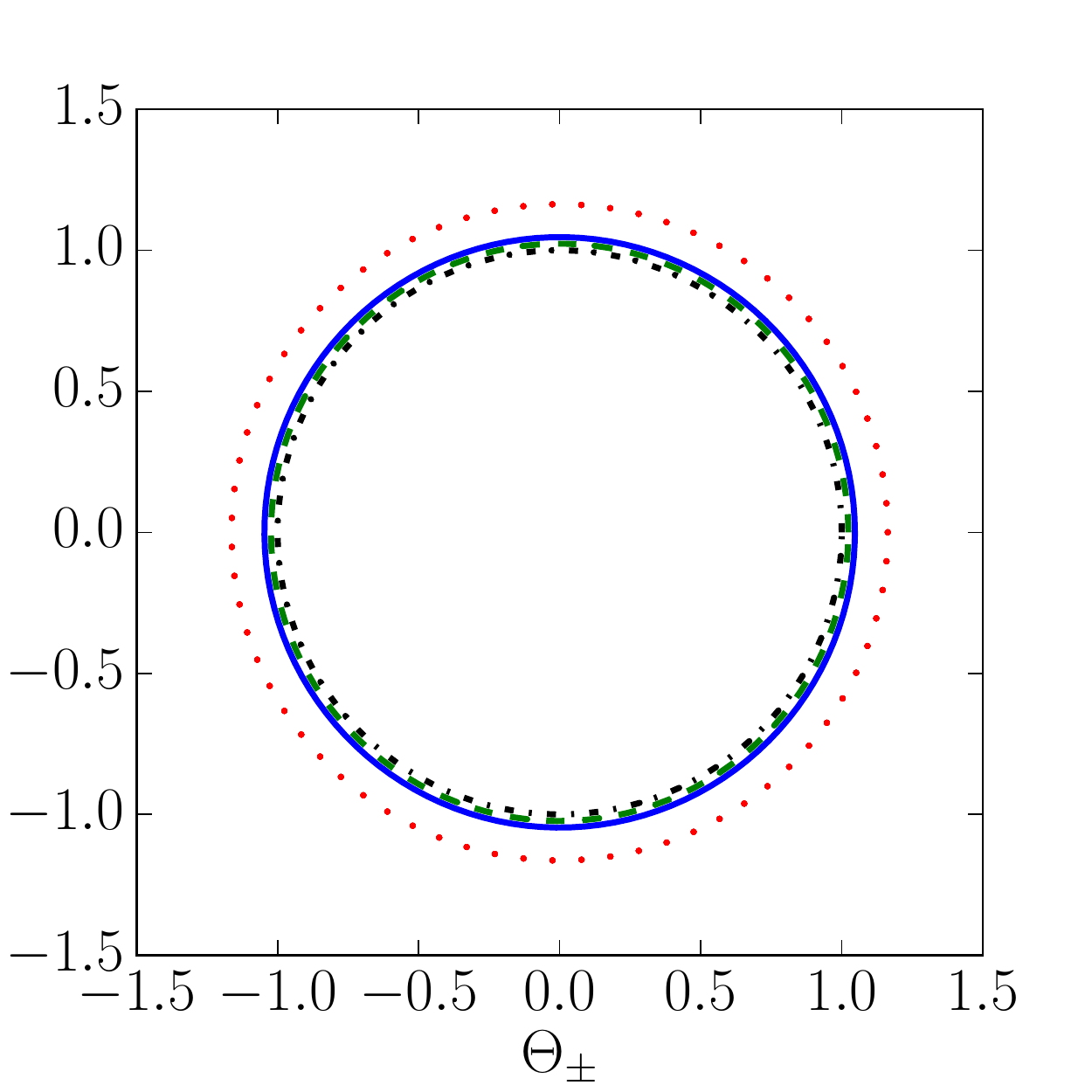}
\includegraphics[width=0.5\textwidth]{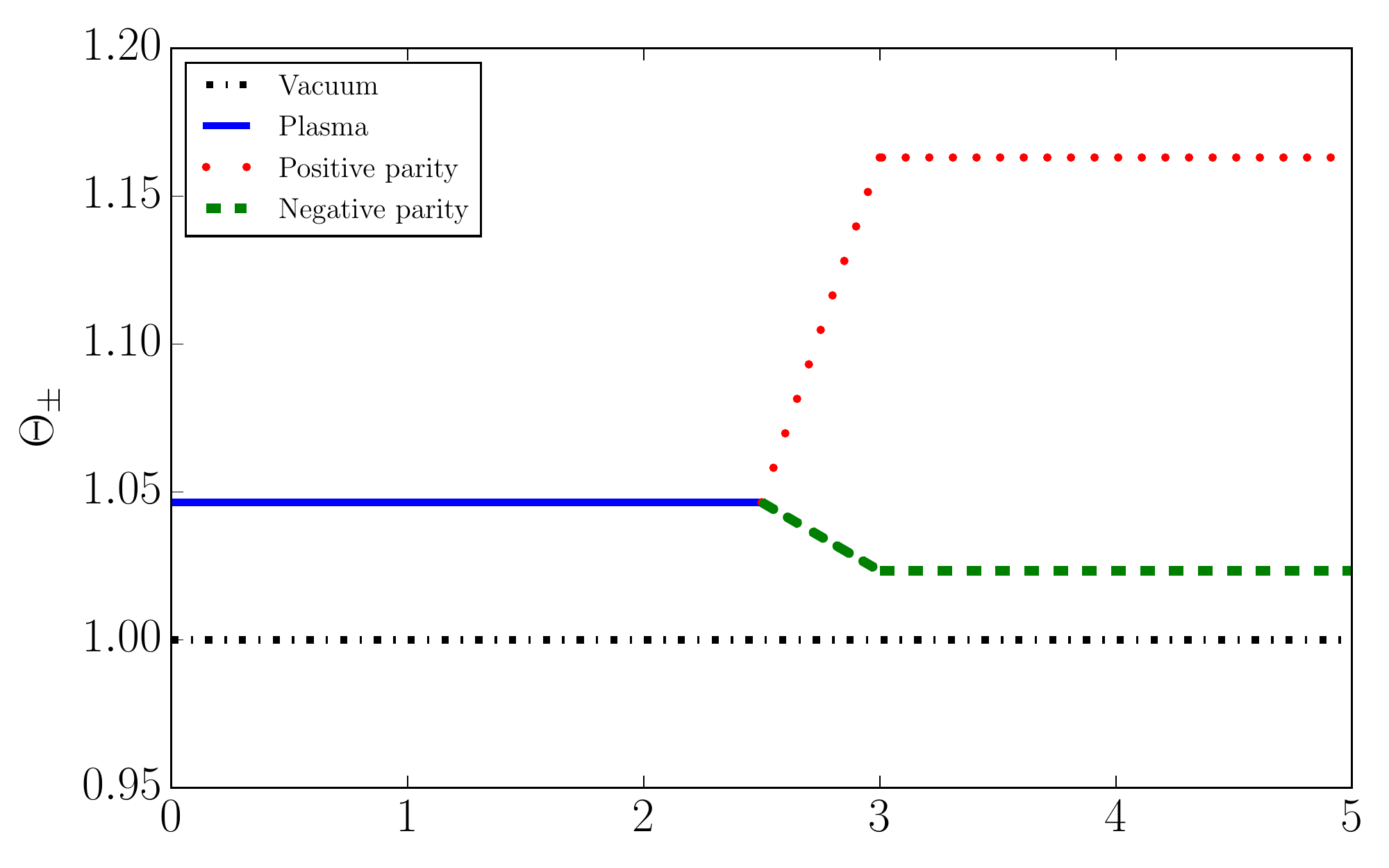}
\caption{\label{ER} Einstein ring for the plasma frequency 
$\omega_{\rm 0} = 0.4\,\omega$ and the cyclotron frequency 
$\omega_{\rm c}= 0.6\,\omega$. The black line corresponds 
to the vacuum $\Theta_{\rm E}$, the blue line is for the 
plasma $\Theta$, and the dashed red and green lines are 
for $\Theta_{+}$ and $\Theta_{-}$ at $\psi = 0$.}
\end{figure}

\begin{figure}
\includegraphics[width=0.42\textwidth]{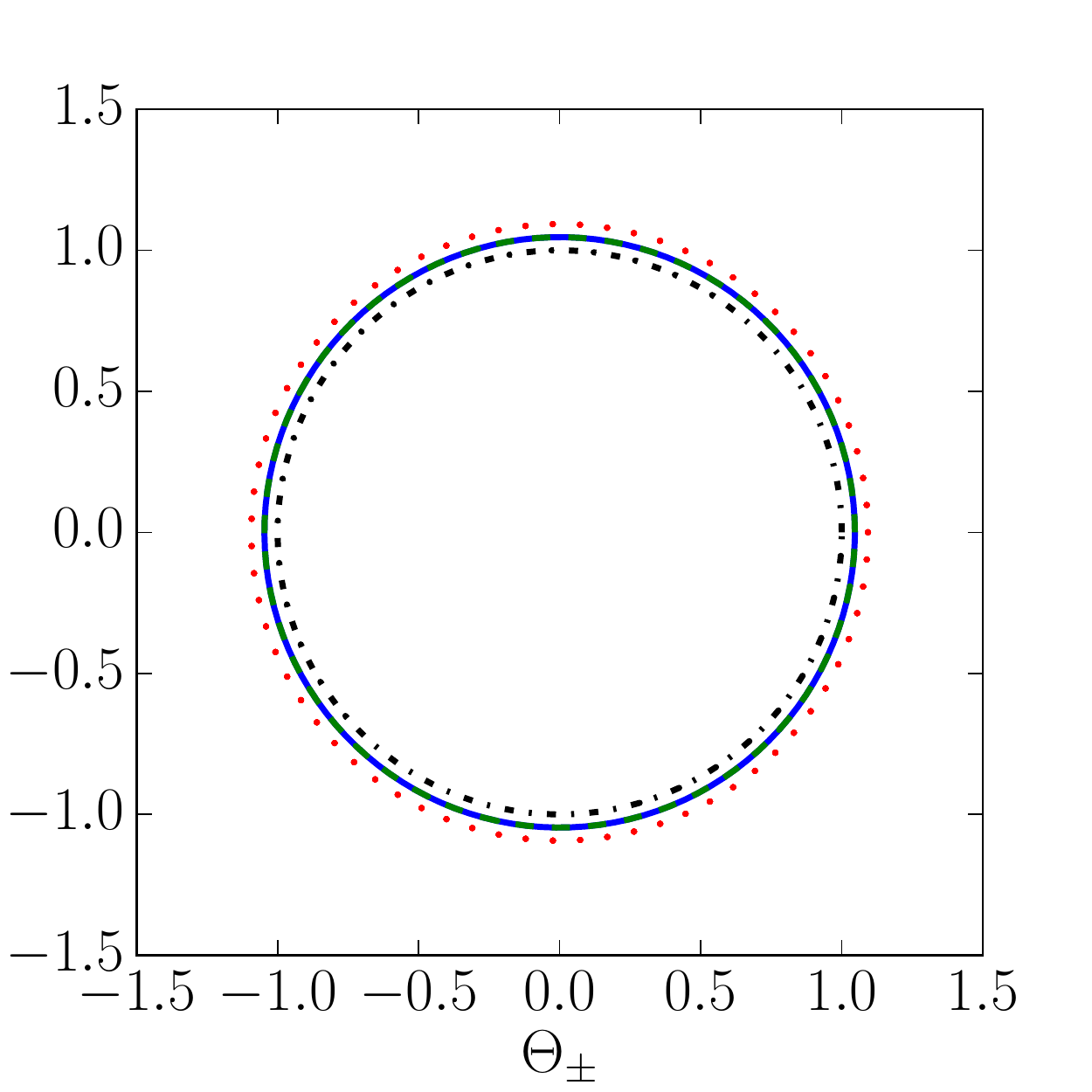}
\includegraphics[width=0.5\textwidth]{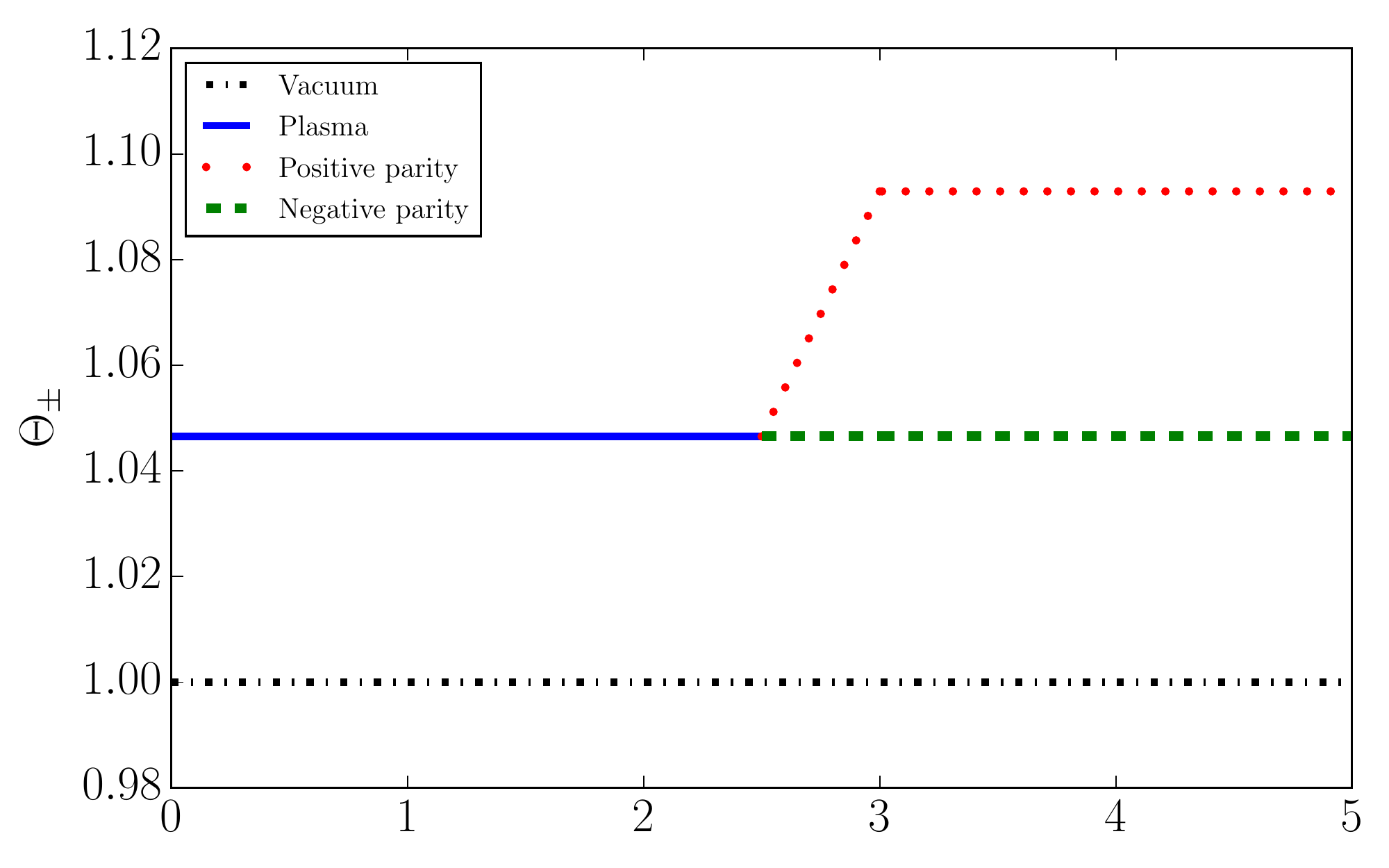}
\caption{\label{ER} Einstein ring for the plasma frequency 
$\omega_{\rm 0} = 0.4\,\omega$ and the cyclotron frequency 
$\omega_{\rm c}= 0.6\,\omega$. The black line corresponds 
to the vacuum $\Theta_{\rm E}$, the blue line is for the 
plasma $\Theta$, and the dashed red and green lines are 
for $\Theta_{+}$ and $\Theta_{-}$ at $\psi = \pi/2$.}
\end{figure}

\begin{figure}
\includegraphics[width=0.42\textwidth]{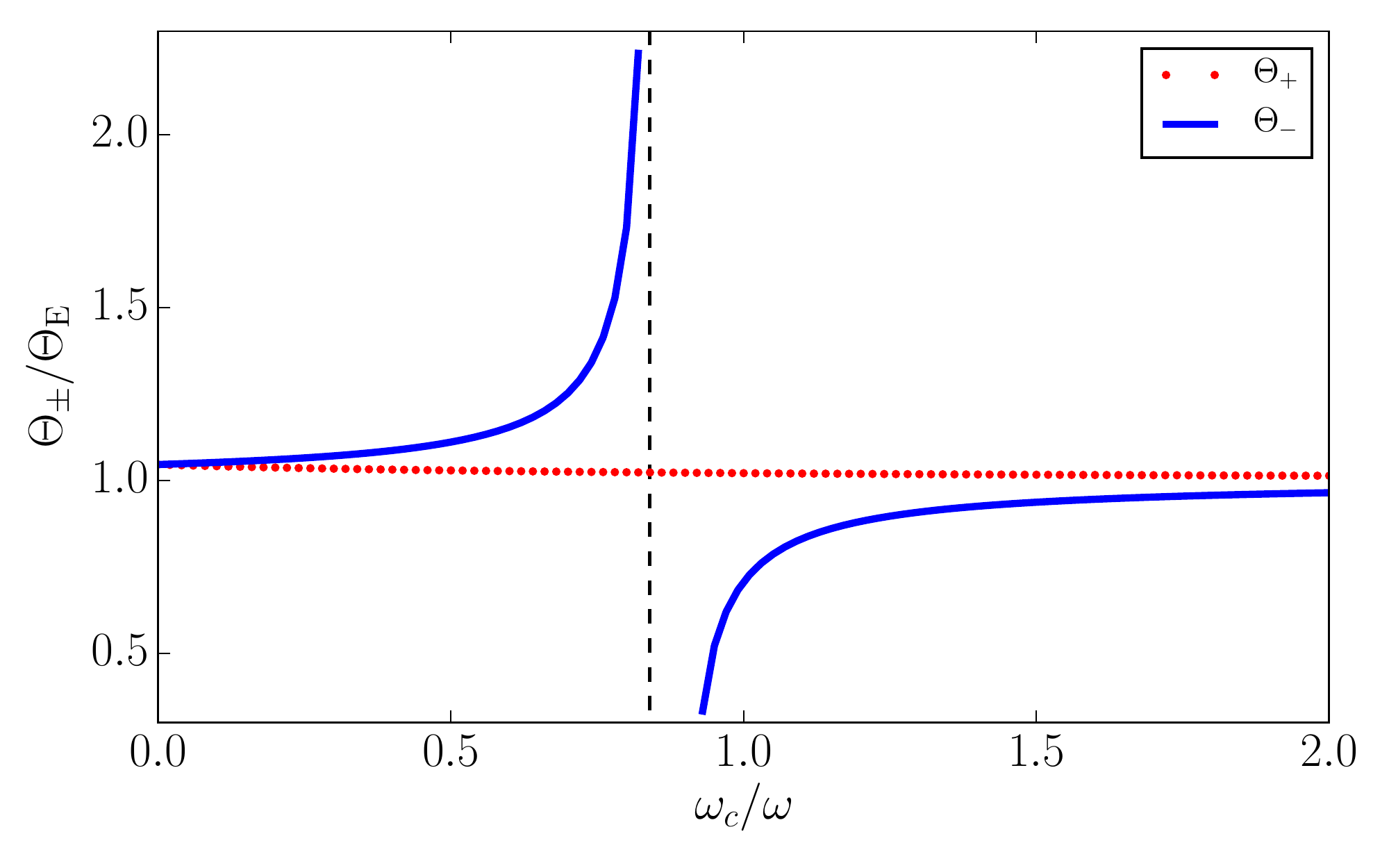}
\includegraphics[width=0.42\textwidth]{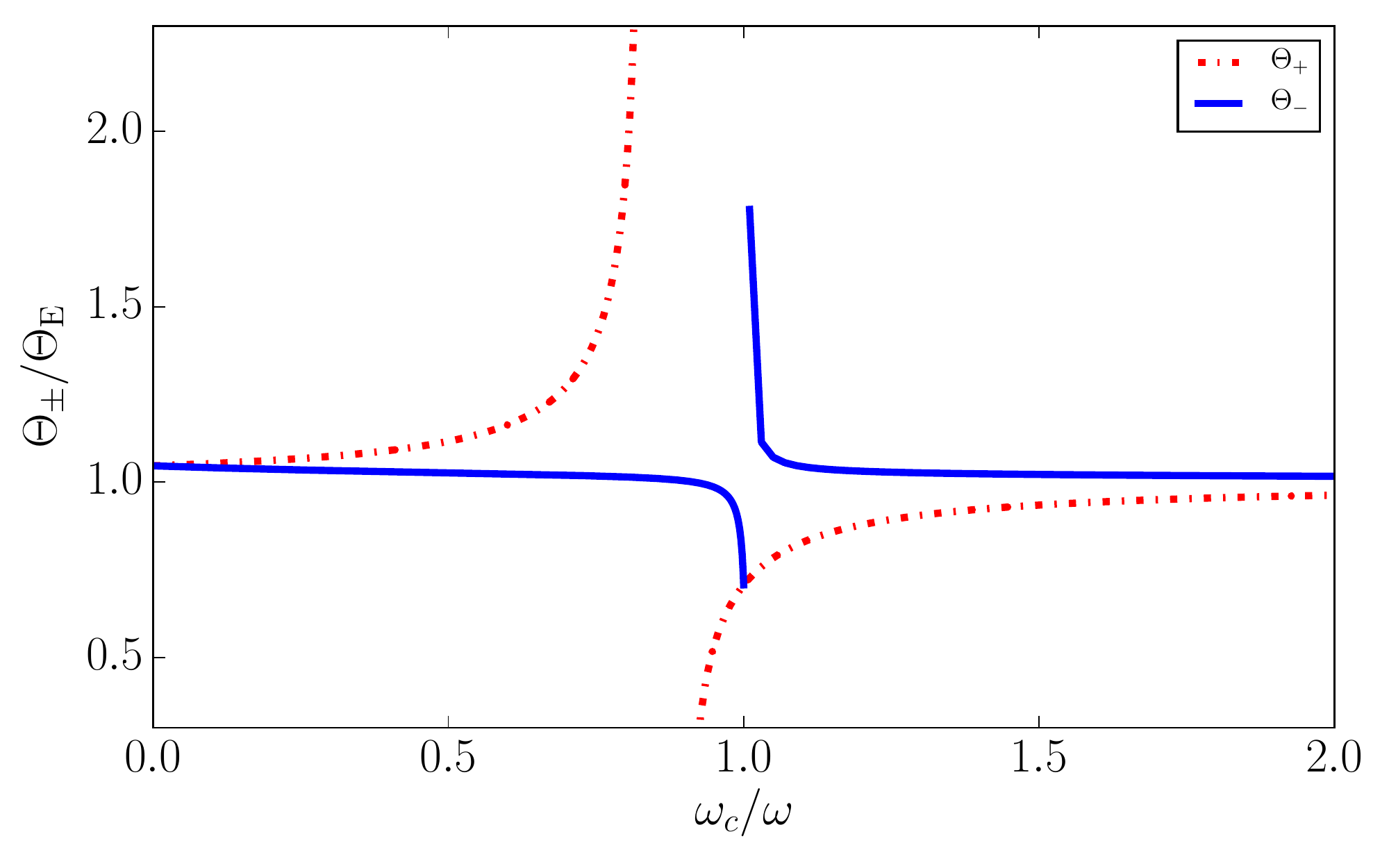}
\caption{Dependence of $\Theta_{\pm}$ on $\omega_c/\omega$ 
for $\omega_0 = 0.4 \omega$ and $\omega_c = 0.6 \omega$ when $\psi=0$ (left panel)
and $\psi=\pi/2$ (right panel). \label{Thetawc}}
\end{figure}

Figure~\ref{Thetawc} shows the angular size of the Einstein ring as a function of the cyclotron frequency. From Fig.~\ref{Thetawc}, we can see that in the resonance state, when $\omega_c\sim 0.84 \omega$, the size of the Einstein ring increases. An observation of the ring in the corresponding range of frequency would detect the change of the size and the form of the Einstein ring due to the existence of a magnetic field. 

Let us now consider the image magnification due to lensing. First, we write $\theta$ in terms of $\beta$. The solution of Eq.~(\ref{beta}) is
\begin{eqnarray}\label{thetotbet}
\theta = \frac{1}{2}\left(\beta \pm 
\sqrt{\beta^2 + 4\Theta_\pm ^{2}}\right)\ .
\end{eqnarray}
We define the image magnification as
\begin{eqnarray}
\mu &=& \bigg\vert\frac{\theta}{\beta}
\frac{d\theta}{d\beta}\bigg\vert\ .
\end{eqnarray}
Using Eq.~(\ref{thetotbet}), we can easily find the following expressions 
\begin{eqnarray}
\mu_{1}^{\pm} &=&
\frac{1}{4}\left[\frac{\beta}{\sqrt{\beta^2 
+ 4\Theta_\pm ^{2}}}+\frac{\sqrt{\beta^2 
+ 4\Theta_\pm ^{2}}}{\beta} + 2\right]\ ,\label{muu1}
\\
\mu_{2}^{\pm} &=&
\frac{1}{4}\left[\frac{\beta}{\sqrt{\beta^2 
+ 4\Theta_\pm ^{2}}}+\frac{\sqrt{\beta^2 
+ 4\Theta_\pm ^{2}}}{\beta} - 2\right]\ .\label{muu2}
\end{eqnarray}
The total magnification is 
\begin{eqnarray}
\mu_{\rm T}^\pm = \mu_{1}^\pm + \mu_{2}^\pm 
&=&
\frac{\beta^2+2\Theta_\pm ^{2}}{\beta\sqrt{\beta^2 + 4\Theta_\pm ^{2}}}.
\end{eqnarray}
The ratio of the two magnifications is
\begin{eqnarray}
\nu^{\pm} = \frac{\mu_{1}^{\pm}}{\mu_{2}^{\pm}} 
&=& \left(\frac{\theta_{1}^{\pm}}{\theta_{2}^{\pm}}\right)^2 = 
\left[\frac{\sqrt{\beta^2 + 4\Theta_\pm ^{2}}+\beta}{\sqrt{\beta^2 + 4\Theta_\pm ^{2}}-\beta}\right]^2\ , 
\end{eqnarray}
where we have used following notations 
\begin{eqnarray}\label{Notation}
&&
\theta_{1}^{\pm}\,\theta_{2}^{\pm} = - \Theta_{\pm}^2,
\quad 
\theta_{1}^{\pm} + \theta_{2}^{\pm} = \beta,
\\\nonumber
&&
\theta_{1}^{\pm} - \theta_{2}^{\pm} = \sqrt{\beta^2+4
\Theta_{\pm}^2}\ .
\end{eqnarray}

\begin{figure}
\includegraphics[width=0.3\textwidth]{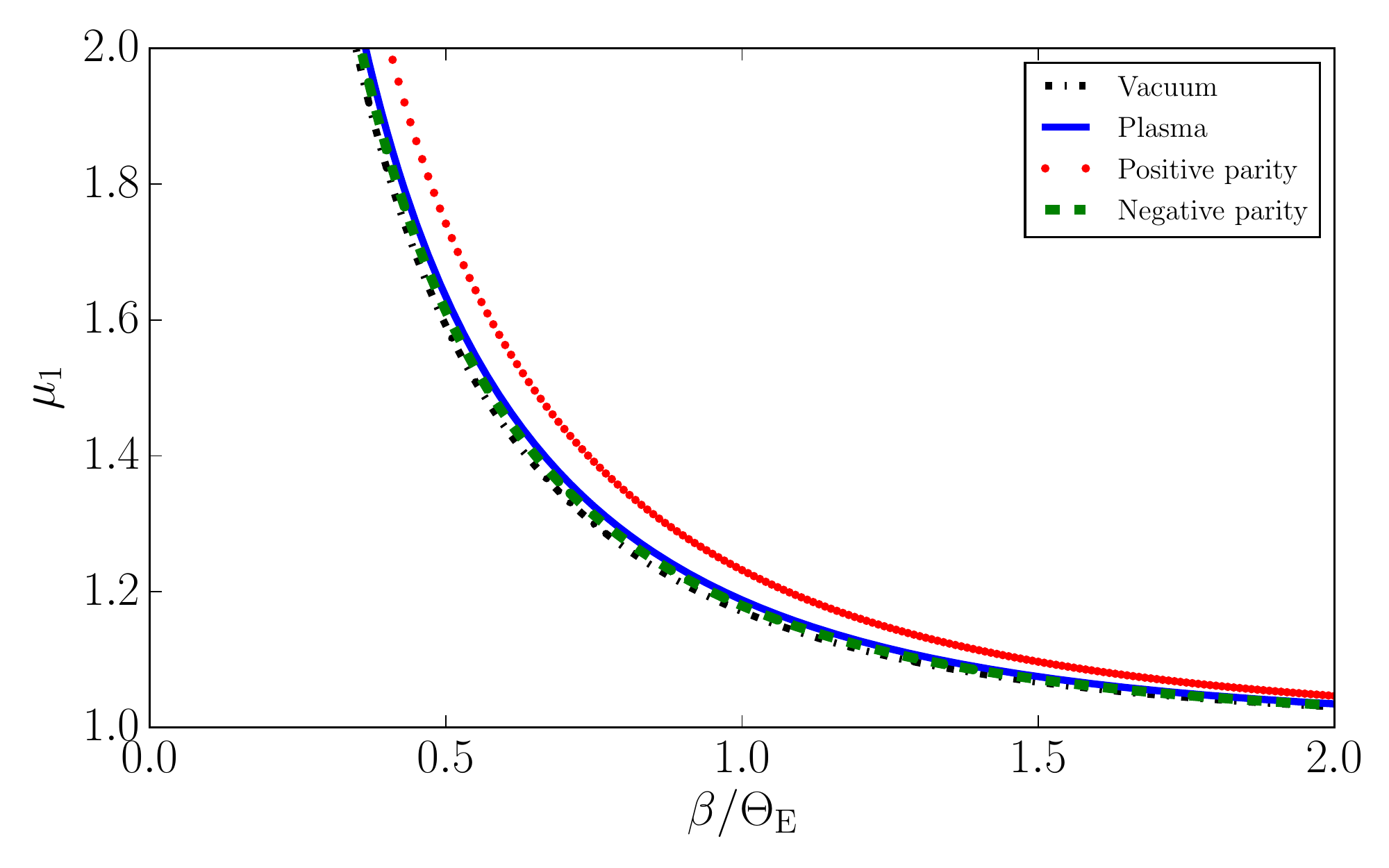}
\includegraphics[width=0.3\textwidth]{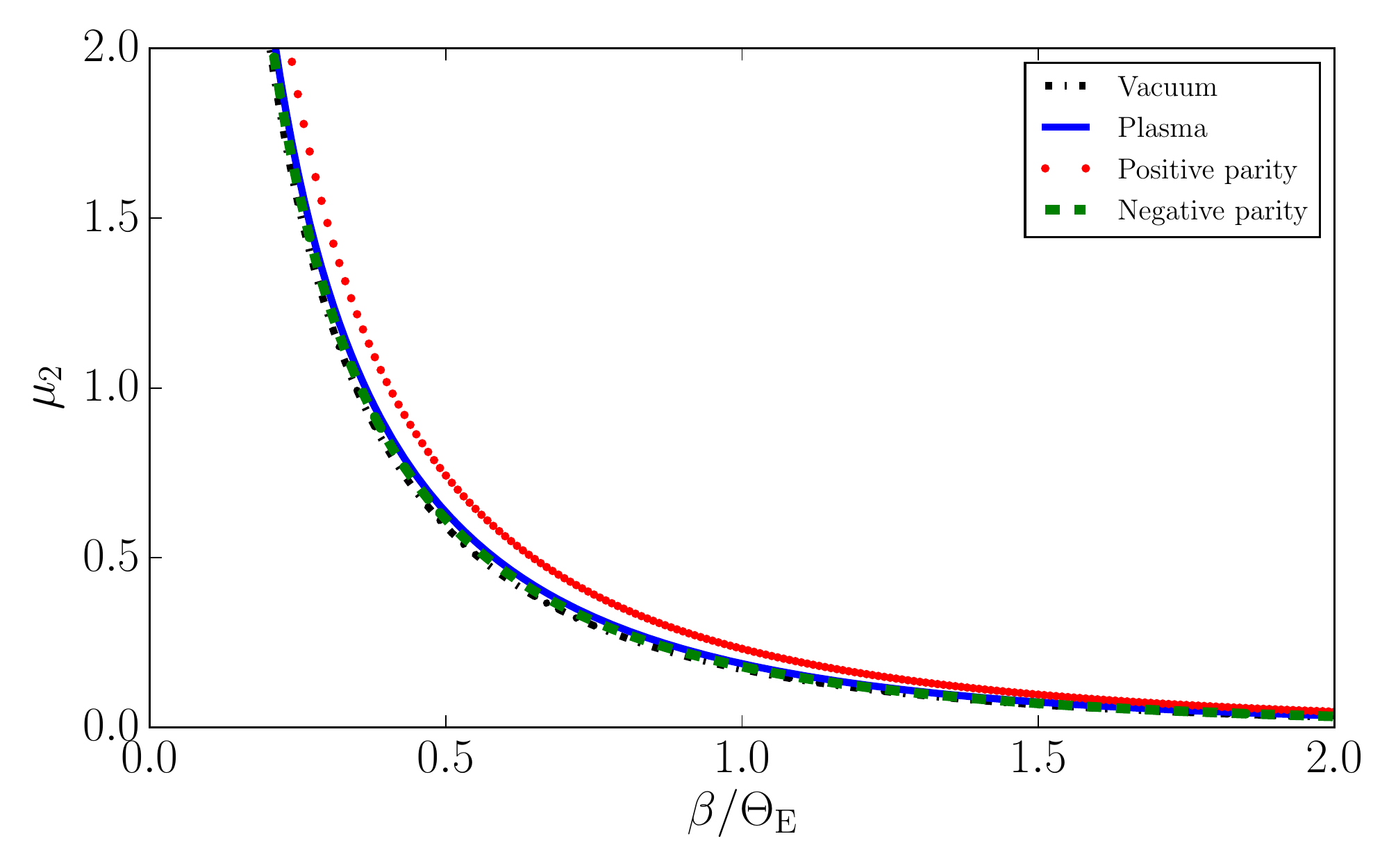}
\includegraphics[width=0.3\textwidth]{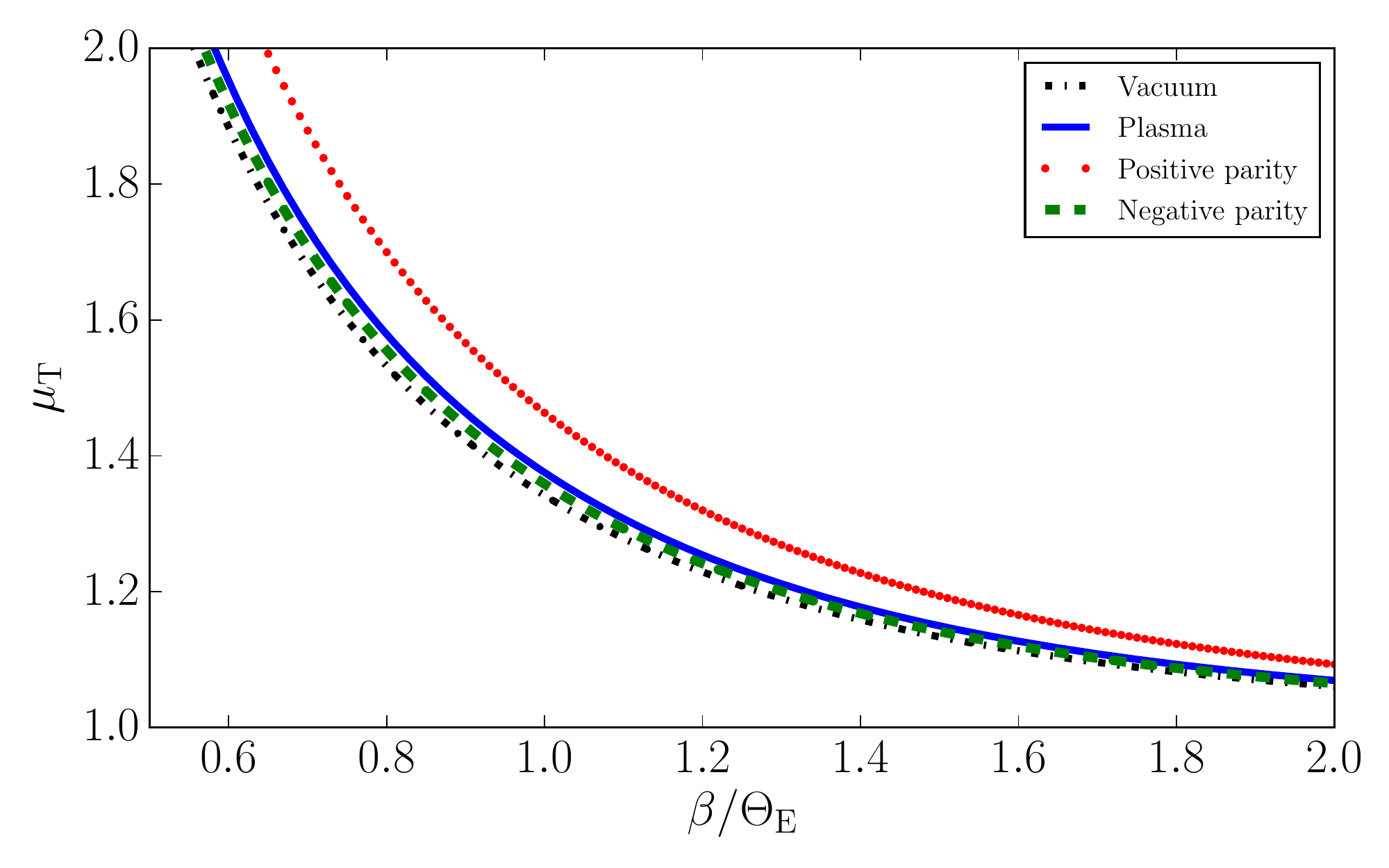}
\caption{\label{magnification1} Magnification for the first image $\mu_1 $ as a function of $\beta/\Theta_{\rm E}$. All lines are plotted for $\omega_0 = 0.4 \omega$ and $\omega_c = 0.6 \omega$ when $\psi=0$ .}
\end{figure}

\begin{figure}
\includegraphics[width=0.3\textwidth]{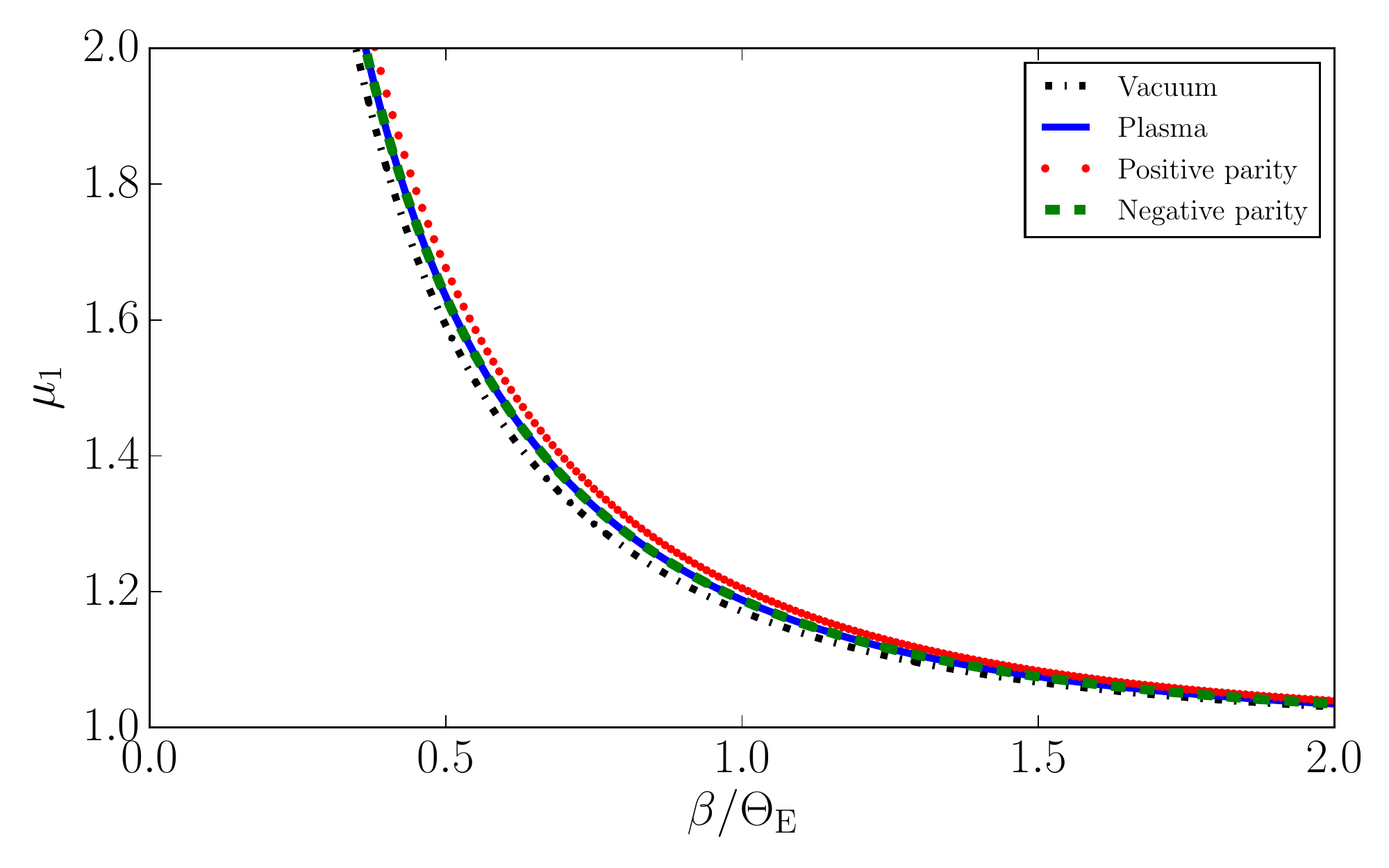}
\includegraphics[width=0.3\textwidth]{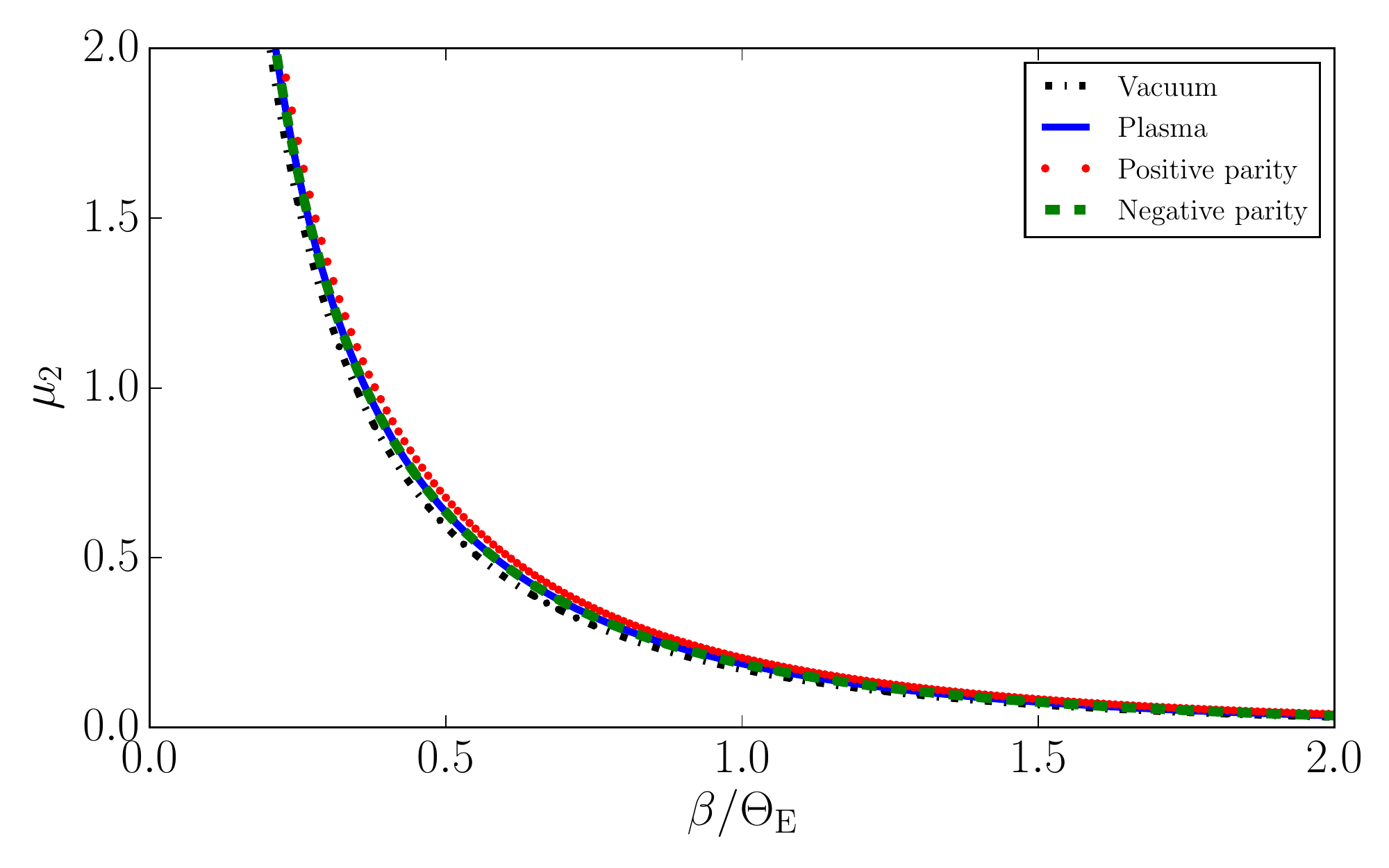}
\includegraphics[width=0.3\textwidth]{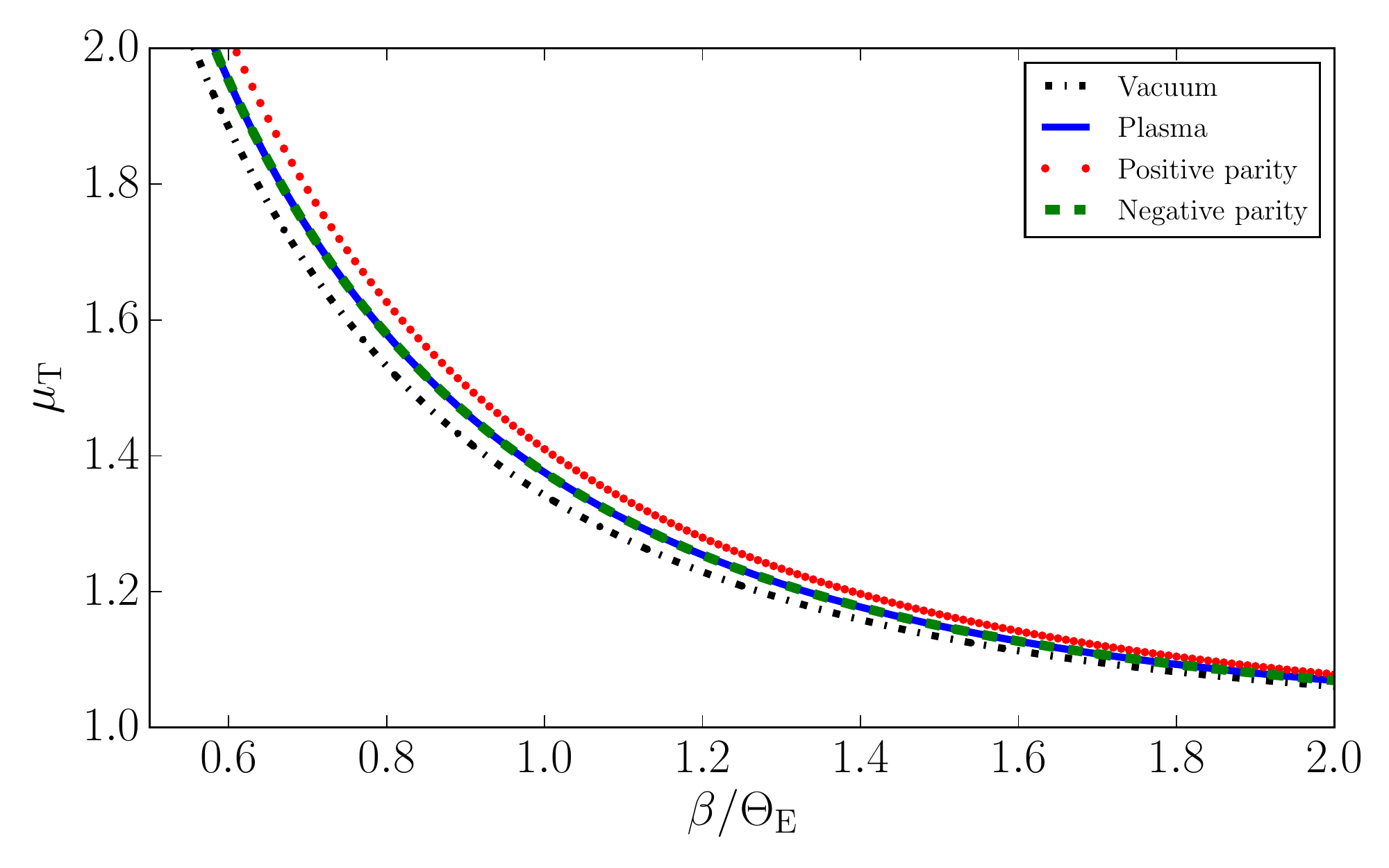}
\caption{\label{magnification2} Magnification for the 
first image $\mu_1 $ as a function of $\beta/\Theta_{\rm E}$. 
All lines are plotted for $\omega_0 = 0.4 \omega$ and 
$\omega_c = 0.6 \omega$ when $\psi=\pi/2$.}
\end{figure}

\begin{figure}
\includegraphics[width=0.49\textwidth]{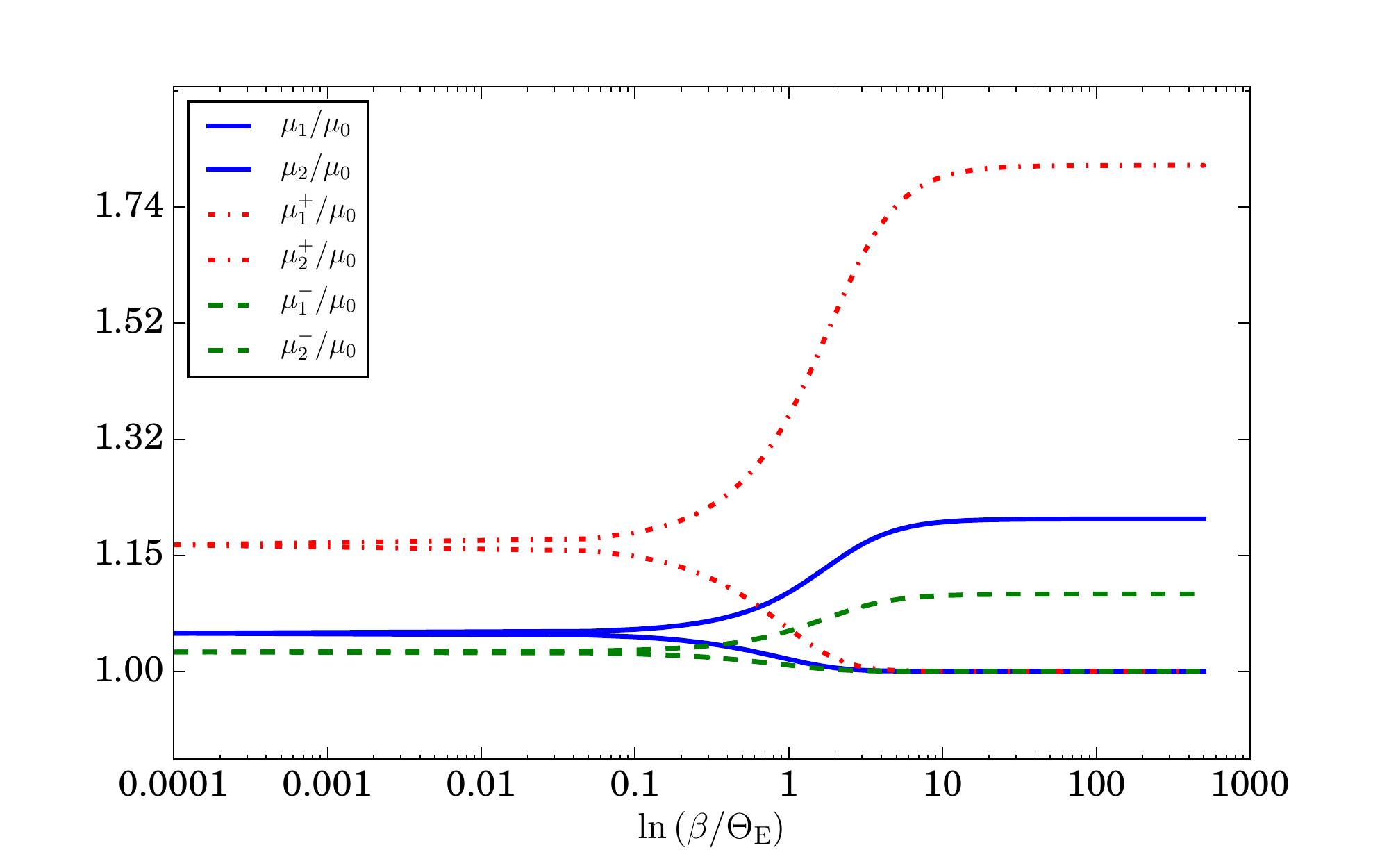}
\includegraphics[width=0.49\textwidth]{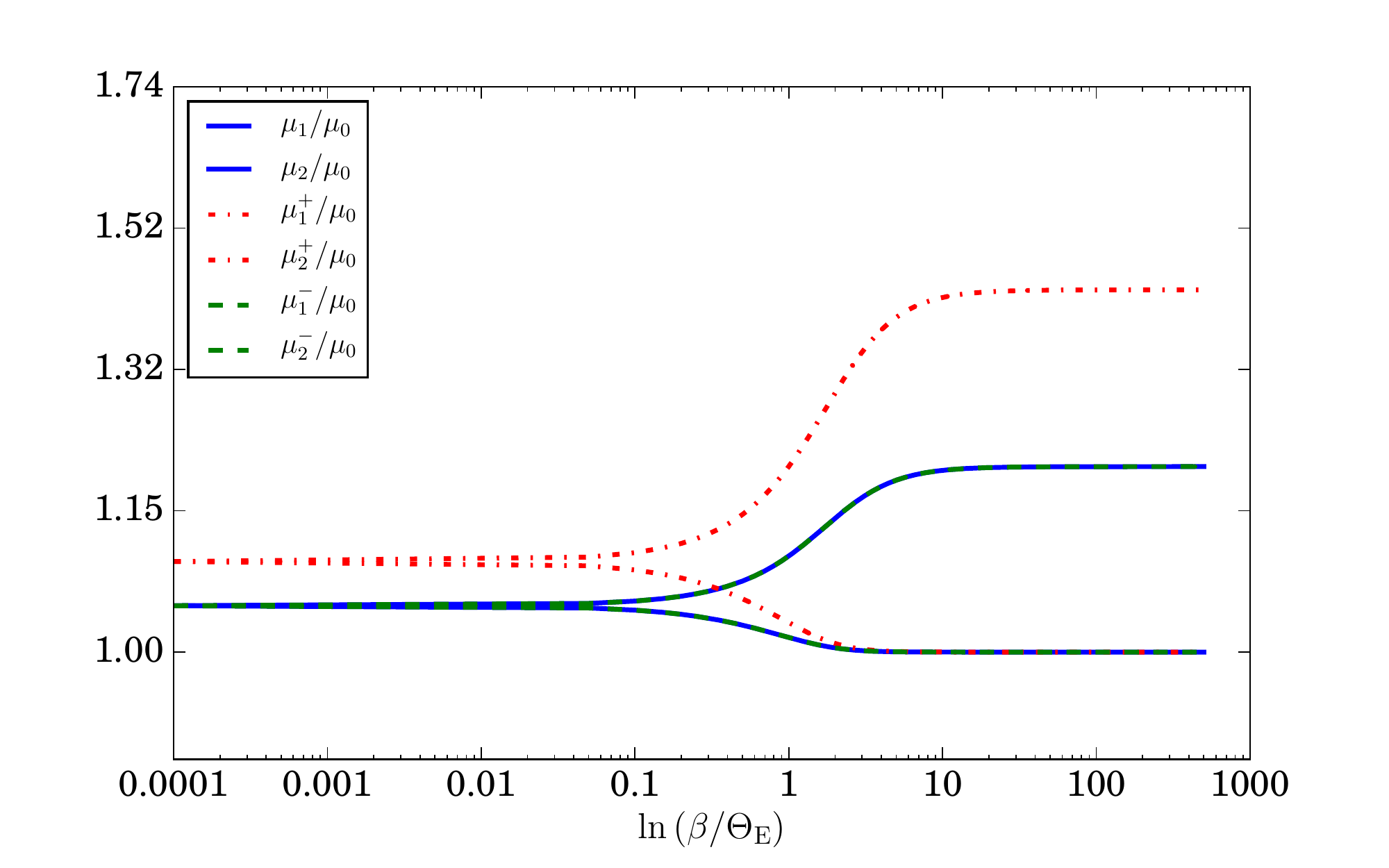}
\caption{\label{splittt} Ratio of the magnification 
$ \mu_{\pm, i}/\mu_0$,  ($i=1,2$) as a function of 
$\beta/\Theta_{\rm E}$. All lines are plotted for 
$\omega_0 = 0.4 \omega$ and $\omega_c = 0.6 \omega$ when $\psi = 0$
(left panel) and $\psi = \pi/2$ (right panel).}
\end{figure}

Figures~\ref{magnification1} and~\ref{magnification2} show the magnification of the first and of the second images due to weak lensing in the presence of a homogeneous plasma and a magnetic field. The upper and lower plots correspond, respectively, to the first and to the second solution for the magnification presented in Eqs.~(\ref{muu1}) and (\ref{muu2}). From these plots we can easily see that due to the magnetic ``Zeeman effect'' the magnification plots split into two lines (dashed and dotted lines) with respect to the unmagnetized plasma case (solid line). Magnetic fields, in principle, cause the amplification of the magnification of image sources (see Fig.~\ref{magnification1} and Fig.~\ref{magnification2}). 

Fig.~\ref{splittt} shows the ratio of different magnifications of the image source; the upper and the lower ``solid'' lines correspond to the case of a plasma without magnetic field, which is the situation 
shown in~\cite{Kogan10}; the dashed and dotted-dashed lines show the split of the line due to the presence of a homogeneous plasma and to the magnetic ``Zeemann'' effect.

A variable source behind a lensing object produces an observable variable image. However, the source and the image will not necessarily vary simultaneously: in general, there will be a time delay between the two events and there are two contributions. First, there is a purely geometrical time delay. Second, there is a delay due to the potential of the lensing object, the so-called Shapiro time delay. 

If the set-up is that illustrated in Fig.~\ref{GL}, we have the following relation
\begin{eqnarray}
D_{\rm l} +D_{\rm ls}-D_{\rm s} = \frac{D_{\rm l}
D_{\rm s}}{2D_{\rm ls}}(\theta - \beta)^2 = c \Delta t_g\ ,
\end{eqnarray} 
where $\Delta t_g$ is the time delay caused by the spacetime geometry.
There are two values of $\theta$ corresponding to the two values of the geometrical time delay. The time delay of one of the images with respect to the other one is 
\begin{eqnarray}\label{tdel1}
\Delta t_g^{\pm} &=& \frac{D_{\rm l}
D_{\rm s}}{2D_{\rm ls}}\left[(\theta_1^{\pm} - \beta)^2-
(\theta_2^{\pm} - \beta)^2\right]\ .
\end{eqnarray} 
Recalling the lens equation in Eq.~(\ref{beta}) and the notations in Eq.~(\ref{Notation}), we can rewrite Eq.~(\ref{tdel1}) as
\begin{eqnarray}
\Delta t_g^{\pm} &=& \frac{D_{\rm l}
D_{\rm s}}{2D_{\rm ls}}\Theta_{\pm}^4
\left[\frac{1}{\theta_1^{\pm 2}}
-
\frac{1}{\theta_2^{\pm 2}} \right]\nonumber
\\\nonumber
&=& - 2M \frac{\beta\sqrt{\beta^2+4
\Theta_{\pm}^2}}{\Theta_{\rm E}^2}\ .
\end{eqnarray} 
The time delay caused by the gravitational potential (Shapiro time delay) is
\begin{eqnarray}
\Delta t_{\rm ls} &=& 
2M \ln\left(\frac{b}{2D_{\rm ls}}\right)\ , 
\\
\Delta t_{\rm l} &=& 
2M\ln\left(\frac{b}{2D_{\rm l}}\right)\ . 
\end{eqnarray} 
The total Shapiro time delay for the gravitational potential can be written as the sum of these two components  
\begin{eqnarray}\label{deltatg}
\Delta t_p &=& \Delta t_{\rm ls}+\Delta t_{\rm l} \nonumber\\
&=& 2M\ln\left(\frac{b}{2D_{\rm ls}}\right)
+
2M\ln\left(\frac{b}{2D_{\rm l}}\right).\ \ 
\end{eqnarray}

Considering that at very large distances from the lens the potential delay is negligible, we can easily 
calculate the difference in the time delay from one of the images to the other one. We consider the distance $D$ such that $D \gg b$ but $D \ll (D_{\rm ls}, D_{\rm l})$. We can rewrite 
Eq.~(\ref{deltatg}) as 
\begin{eqnarray}
\Delta t_p 
&=&2M
\left[\ln\left(\frac{b}{2D}\right)
+
\ln\left(\frac{b}{2D}\right)
+\ln\left(\frac{D^2}{D_{\rm ls}D_{\rm s}}\right)
\right]
\nonumber\\ 
&= &4M
\ln\left(\frac{b}{2D}\right)+2M\ln\left(\frac{D^2}{D_{\rm ls}D_{\rm s}}\right)\ .
\end{eqnarray}
If we write the impact parameter as $b=\theta D_{\rm l}$, we can find the difference of two time delays
\begin{eqnarray}
\Delta t_p^{\pm} 
&=& 4M \left[\ln\left(\frac{b_1}{2D}\right)
-
\ln\left(\frac{b_2}{2D}\right)\right]
\\\nonumber
&=& 4M \ln\left(\frac{\theta_1}{\theta_2}\right)\ .
\end{eqnarray}

Lastly, the total time delay that arises from both the geometry and the gravitational potential turns out to be
\begin{eqnarray}
\Delta T_{\pm} 
&=& 4M
\ln\left[\frac{\sqrt{\beta^2 + 4\Theta_\pm ^{2}}+\beta}{\sqrt{\beta^2 + 4\Theta_\pm ^{2}}-\beta}\right]
-
 2M \frac{\beta\sqrt{\beta^2+4
\Theta_{\pm}^2}}{\Theta_{\rm E}^2}\ .
\end{eqnarray}

\begin{figure}
\includegraphics[width=0.42\textwidth]{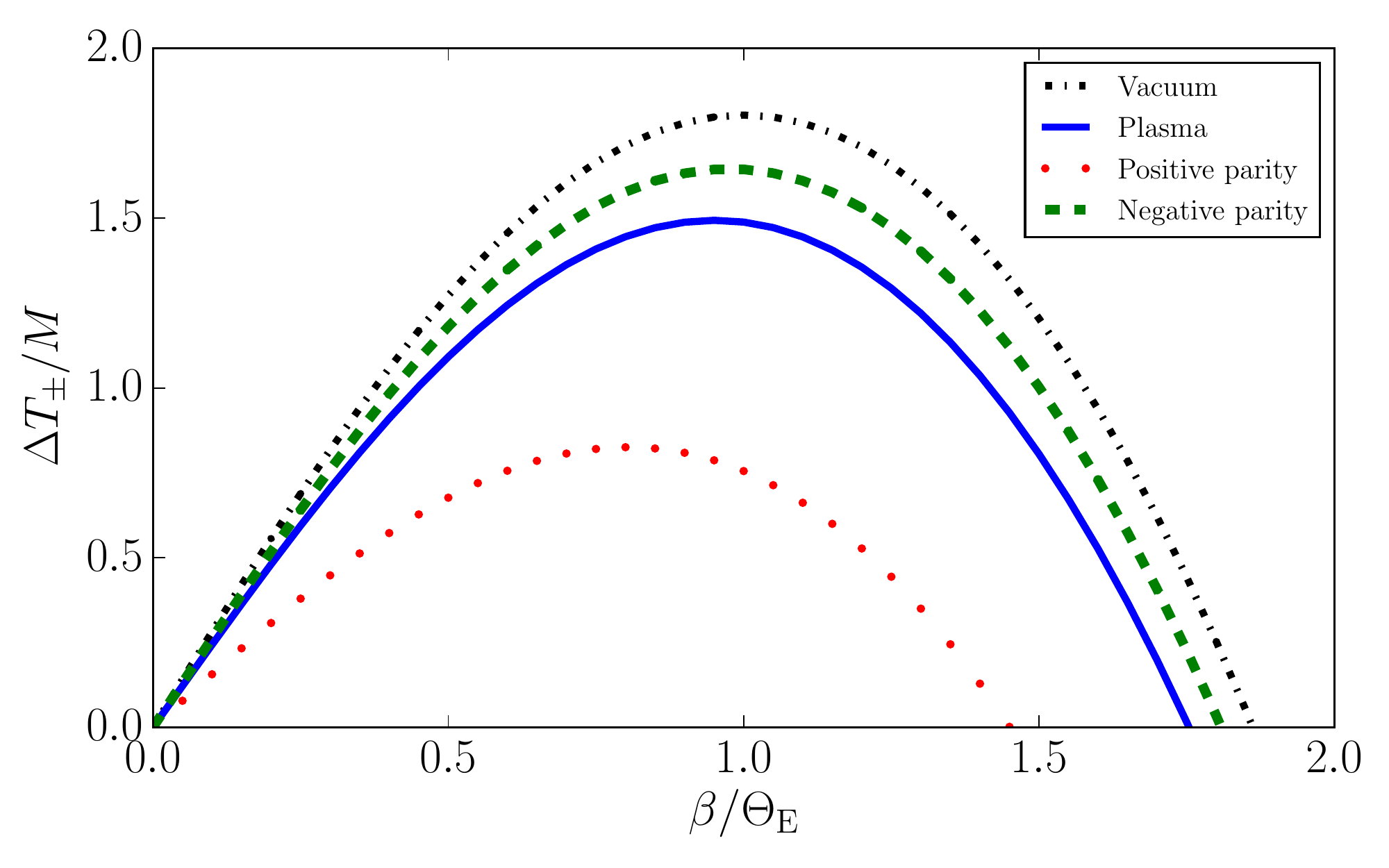}
\includegraphics[width=0.42\textwidth]{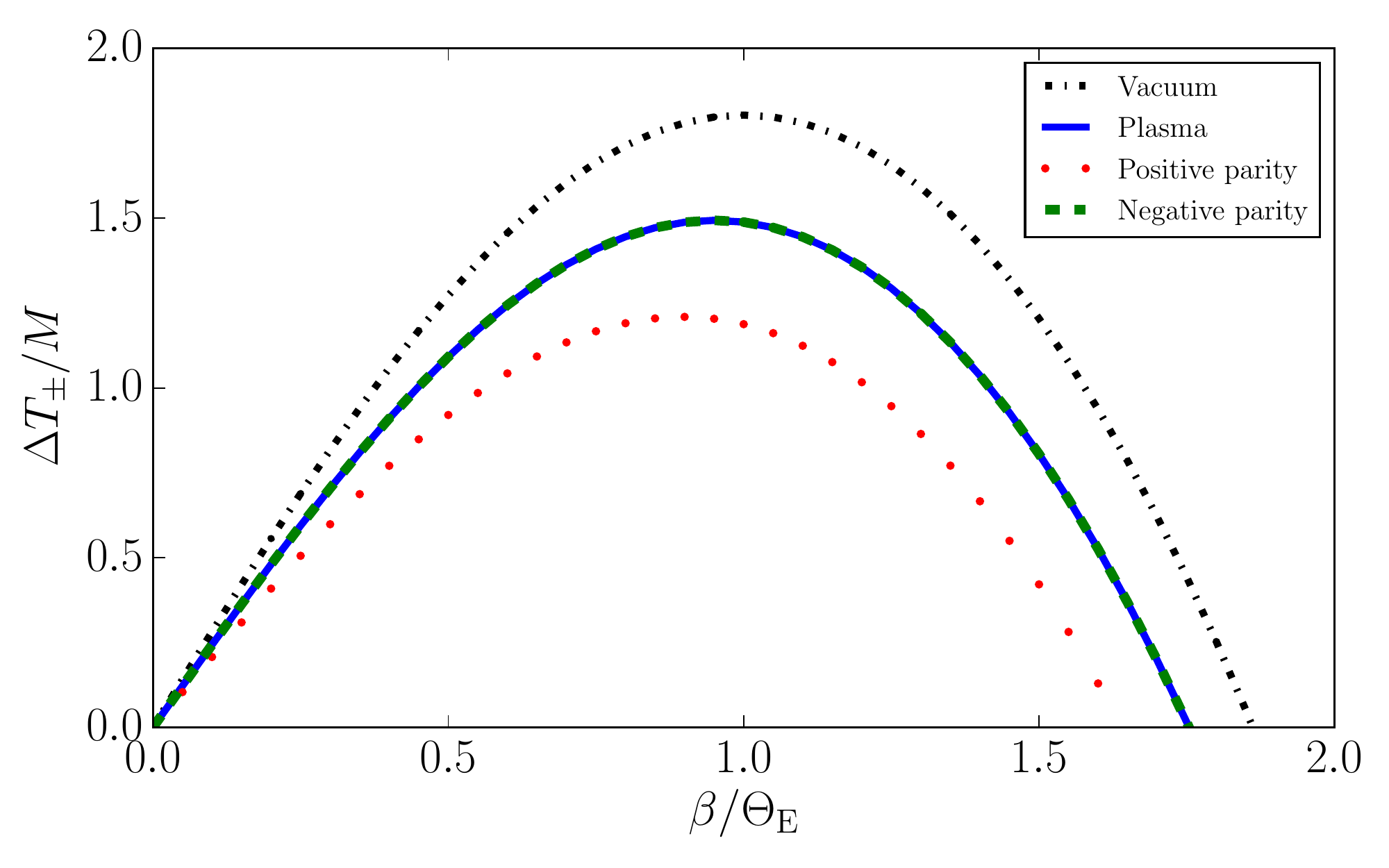}
\caption{\label{timedelay} Time delay 
$\Delta T^{\pm}$ as a function of $\beta/\Theta_{\rm E}$ 
when $\psi = 0$ (left panel) and $\psi = \pi/2$ (right panel).}
\end{figure}

Fig.~\ref{timedelay} shows the dependence of the time delay on the angle $\beta$. As we can see from this plot, the presence of the plasma and of the magnetic field causes a shift of the peak of the time delay. Moreover, the positive and negative time delays due to the magnetic split (Zeeman effect) have their maximum at different values of $\beta$. Fig.~\ref{timedelay} also shows the time delay in the case of vacuum (dot-dashed line in Fig.~\ref{timedelay}).  

\subsection{Inhomogeneous plasma}

Let us now study the effects of an inhomogeneous magnetized plasma. To do this, we assume that $h=1$ in the expression for the plasma frequency. This can be regarded as a toy model for a preliminary study. Substituting Eq.~(\ref{Dangle}) into Eq.~(\ref{Maineq}), we obtain the following expression for $\beta$
\begin{eqnarray}\label{betain}
\beta &=&\theta - 
\frac{\Theta_{\pm}^2}{\theta}
-\frac{\Phi_{\pm}}{\theta^2} \ ,
\end{eqnarray}
where 
$$
\Phi_{\pm} = \frac{\pi \,\Theta_{\rm E}^2}{4}\,
\frac{R_0^2}{2M\, D_{\rm l}}
\,\frac{\omega_0^2}{\omega^2}\,\left(1-\frac{\omega^2_0}{\omega^2}
-\frac{\omega^2_0}{\omega^2}\frac{\omega_c}{\omega}
\,f_{\pm}\left(\omega_c,\omega_0\right)\right)^{-1}\ .
$$
The brightness magnification of the source can be calculated through the formula
\begin{eqnarray}
\mu = \sum_{k}^N\bigg\vert\frac{\theta_k}{\beta} 
\frac{d \theta_k}{d \beta}\bigg\vert\ ,
\end{eqnarray}
where $N$ is the number of images of the source (star).
In this case, the lens equation can be written as 
\begin{eqnarray}\label{Lenseq}
\theta^3 - \beta \theta^2 -\Theta_{\pm}^2\theta
- \Phi_{\pm} = 0  \ .
\end{eqnarray}
In order to solve the equation, we introduce the new variable $x = \theta - \frac{\beta}{3}$, which we plug into Eq.~(\ref{Lenseq}). We get
\begin{eqnarray}\label{Eqx}
x^3 + p\,x^2 +q = 0  \ ,
\end{eqnarray}
where
\begin{eqnarray}
p &=& - \frac{1}{3}\beta^2 - \Theta_{\pm}^2 \ ,
\\
q &=& -\frac{2}{27}\beta^3 - \frac{1}{3}\beta 
\Theta_{\pm}^2  - \Phi_{\pm}\ . 
\end{eqnarray}
Note  that Eq.~(\ref{Eqx}) has three real solutions 
only in case the following condition holds
$$
\frac{q^2}{4} + \frac{p^3}{27} = 0 \ ,
$$ 
and those solutions have the form
\begin{eqnarray}\label{xk}
x_k = 2\sqrt[3]{s} \cos \left(
\frac{\gamma+2\pi k}{3}\right), \qquad k= 0,\,1,\,2\ ,
\end{eqnarray}
where 
$$
s = \sqrt{-\frac{p^3}{27}}\,, \qquad  
\cos\gamma = - \frac{q}{2 r}\ . 
$$

The magnification for the gravitational lens surrounded 
by an inhomogeneous magnetized plasma assumes the form
\begin{eqnarray}\label{Mu}
\mu &=& \sum_{k}^N\bigg\vert\frac{\theta_k}{\beta} 
\frac{d \theta_k}{d \beta}\bigg\vert 
=
\sum_{k}^N 
\bigg\vert \left(\frac{x_k}{\beta}+\frac{1}{3}\right) 
\left(\frac{d x_k}{d \beta}+\frac{1}{3}\right)\bigg\vert
\\\nonumber
&=& \sum_{k}^N 
\bigg\vert \frac{1}{3\beta}
\left(2\sqrt[3]{s}\cos\frac{\gamma+2\pi k}{3}+
\frac{\beta}{3}\right)
\\\nonumber
&&
\times\left(1 - 2\sqrt[3]{s}\, \frac{d\gamma}{d\beta}
\sin\frac{\gamma+2\pi k}{3}
+ \frac{2}{\sqrt[3]{s^2}}\frac{ds}{d\beta}
\cos\frac{\gamma+2\pi k}{3}\right)\bigg\vert. 
\end{eqnarray}

It is now easy to calculate the Einstein ring in the 
case of an inhomogeneous plasma by setting $\beta = 0$ 
in Eq.~(\ref{xk}). The tree  different values of the Einstein ring are
%
\begin{eqnarray}
\label{T1}
\Theta_{1,\pm}^p &=& 
\frac{2}{\sqrt{3}}\Theta_{\pm} 
\cos\left[\frac{1}{3}\arccos
\left(\frac{3\sqrt{3}}
{2}\frac{\Phi_{\pm}}
{\Theta_{\pm}^3}\right)\right] ,\ \ 
\\
\label{T2}
\Theta_{2,\pm}^p &=&
\frac{2}{\sqrt{3}}\Theta_{\pm} 
\cos\left[\frac{1}{3}\arccos
\left(\frac{3\sqrt{3}}{2}\frac{\Phi_{\pm}}
{\Theta_{\pm}^3}\right)+\frac{2\pi}{3}\right],\ \  
\\
\label{T3}
\Theta_{3,\pm}^p &=& \frac{2}{\sqrt{3}}\Theta_{\pm} 
\cos\left[\frac{1}{3}\arccos
\left(\frac{3\sqrt{3}}{2}\frac{\Phi_{\pm}}
{\Theta_{\pm}^3}\right)+\frac{4\pi}{3}\right].\ \ 
\end{eqnarray}
%

\section{Conclusions \label{conclusion}}

In this paper we have studied the gravitational lensing in the weak field approximation, extending previous work in the literature. We have considered a plasma and a magnetic field around a gravitational source. Our results can be summarized as follows: 

\begin{itemize}

\item In the presence of a magnetic field, we may observe the split of the Einstein ring, as the counterpart of the Zeeman effect. When the cyclotron frequency approaches the plasma frequency, the size and the form of the ring change because of the presence of a resonance state. This is a pure magnetic effect and can potentially help to study magnetic fields through gravitational lensing effects. 

\item We have studied the magnification of the image source due to weak lensing in the presence of a homogeneous plasma and of a magnetic field. Due to the magnetic ``Zeeman effect'', the magnification splits into two additional components with respect to the unmagnetized plasma case. 

\item We have also studied the time delay due to the geometry and the gravitational field around a gravitational source. We found that the presence of a plasma and of a magnetic field sufficiently changes the time delay depending on the angle $\beta$.

\item As a toy model, we have considered a power law density plasma. Inhomogeneities in the plasma also lead to image source magnifications. We found that an inhomogeneous plasma increases the source image magnification. 

\end{itemize}

\section*{Acknowledgments}

This work is supported by the Grants No. VA-FA-F-2-008 of the Uzbekistan Agency for Science and Technology, by the Abdus Salam International Centre for
Theoretical Physics through Grant No. OEA-NT-01 and
the Volkswagen Stiftung, Grant No. 86 866. This research
is partially supported by an Erasmus+ exchange grant
between SU and NUUz. A. A. acknowledges the TWAS
associateship program for support. B. A. and A. A. acknowledge the Faculty of Philosophy
and Science, Silesian University in Opava, Czech Republic,
Inter-University Centre for Astronomy and Astrophysics,
Pune, India, and Goethe University, Frankfurt am Main,
Germany, for warm hospitality.

\bibliographystyle{JHEP}
\bibliography{gravreferences}

\end{document}